\documentclass{aa}  
\usepackage[utf8]{inputenc}
\usepackage{booktabs}

\usepackage{graphicx}
\usepackage{verbatim}
\usepackage{txfonts}

\newcommand{\igr}{IGR~J17511$-$3057}
\newcommand{\nicer}{NICER}
\newcommand{\xmm}{XMM-Newton}
\newcommand{\rxte}{RXTE}
\newcommand{\nustar}{NuSTAR}
\newcommand{\chandra}{Chandra}

\newcommand{\inte}{INTEGRAL}

\DeclareUnicodeCharacter{2212}{-}
\begin{document}

   \title{The 2025 outburst of \igr{}: timing and spectral insights from \nicer{} and \nustar{}}

   \titlerunning{the 2025 outburst of IGR J17511-3057}
    \authorrunning{A.~Sanna et al.}

\author{A.~Sanna\inst{1}
        \and G.~K.~Jaisawal\inst{2}
        \and T.~E.~Strohmayer\inst{3}
        \and G.~Illiano\inst{4}
        \and A.~Riggio\inst{1}
        \and A.~Papitto\inst{5}
        \and T.~Di~Salvo\inst{6}
        \and L.~Burderi\inst{1,7}
        \and J.~B.~Coley\inst{8,9}
        \and D. Altamirano\inst{10}
        \and C. Malacaria\inst{5}
        \and A. Anitra\inst{1,6} 
        \and M.~Ng\inst{11,12}
        \and D.~Chakrabarty\inst{13}
        \and T.~Boztepe\inst{14}
        \and A. C. Albayati\inst{10}
}
\institute{
  Dipartimento di Fisica, Universit\`a degli Studi di Cagliari, SP
Monserrato-Sestu, KM 0.7, Monserrato, 09042 Italy \\
  \email{andrea.sanna@dsf.unica.it}
  \and
  DTU Space, Technical University of Denmark, 
  Elektrovej 327-328, DK-2800 Lyngby, Denmark
  \and
  Astrophysics Science Division and Joint Space-Science Institute, NASA's Goddard Space Flight Center, Greenbelt, MD 20771, USA
  \and
  INAF–Osservatorio Astronomico di Brera, Via Bianchi 46, I-23807, Merate (LC), Italy
  \and
  INAF Osservatorio Astronomico di Roma, Via Frascati 33, 00078 Monte Porzio Catone (RM), Italy
  \and
  Dipartimento di Fisica e Chimica - Emilio Segr\`e, Universit\`a di Palermo, via Archirafi 36 - 90123 Palermo, Italy
  \and
  INAF/IASF Palermo, via Ugo La Malfa 153, I-90146 Palermo, Italy
  \and
  Department of Physics and Astronomy, Howard University, Washington, DC 20059, USA
  \and
  CRESST and NASA Goddard Space Flight Center, Astrophysics Science Division, 8800 Greenbelt Road, Greenbelt, MD
  \and
  School of Physics and Astronomy, University of Southampton, Southampton SO17 1BJ, UK
  \and 
  Department of Physics, McGill University, 3600 rue University, Montr\'eal, QC H3A 2T8, Canada
  \and 
  Trottier Space Institute at McGill University, 3550 rue University, Montr\'eal, QC H3A 2A7, Canada
  \and
  MIT Kavli Institute for Astrophysics and Space Research, Massachusetts Institute of Technology, Cambridge, MA 02139, USA
  \and
  Istanbul University, Graduate School of Sciences, Department of Astronomy and Space Sciences, Beyaz\i t, 34119, \.Istanbul, T\"urkiye
}

   \date{Received; accepted 2025 September 18 }

  \abstract
   {\igr{} is an accreting millisecond X-ray pulsar and a known type-I burster. The source was observed in outburst for the first time in 2009 and again in 2015, after which it started a decade-long quiescence phase.} 
   {The source was observed in a new outburst phase starting in February 2025 and lasting at least nine days. We investigated the spectral and temporal properties of \igr{}, aiming to characterise its current status and highlight possible long-term evolution of its properties.}
   {We analysed the available \nicer{} and \nustar{} observations performed during the latest outburst of the source. We updated the ephemerides of the neutron star and compared them to previous outbursts to investigate its long-term evolution. We also performed spectral analysis of the broadband energy spectrum in different outburst phases, and investigated the time-resolved spectrum of the type-I X-ray burst event observed with \nustar{}.}
   {We detected X-ray pulsations at the frequency of $\sim 245$~Hz. The long-term evolution of the neutron star ephemerides suggests a spin-down derivative of $\sim -2.3\times 10^{-15}$~Hz/s, compatible with a rotation-powered phase while in quiescence. Moreover, the evolution of the orbital period and the time of the ascending node suggests a fast orbital shrinkage, which challenges the standard evolution scenario for this class of pulsars involving angular momentum loss via gravitational wave emission. The spectral analysis revealed a dominant power-law-like Comptonisation component, along with a thermal blackbody component, consistent with a hard state. Weak broad emission residuals around $6.6$~keV suggest the presence of a K$\alpha$ transition of neutral or He-like Fe originating from the inner region of the accretion disc. Self-consistent reflection models confirmed a moderate ionisation of the disc truncated at around (82-370)~km from the neutron star. Finally, the study of the type-I X-ray burst revealed no signature of photospheric radius expansion. We found marginally significant burst oscillations during the rise and decay of the event, consistent with the neutron star spin frequency.}
   {}
   {}


   \maketitle
%

\section{Introduction}

Accreting millisecond X-ray pulsars (AMXPs) are neutron stars (NSs) in low-mass X-ray binaries (LMXBs) that have been spun up by sustained mass transfer from a sub-solar companion star \citep{Alpar82, Wijnands:1998vk}. LMXBs undergo Roche-lobe overflow from a companion with less than one solar mass. Because the material streaming through the inner Lagrange point carries angular momentum, it forms an accretion disc around the NS, gradually spinning it up to several hundred Hz rotational frequencies as the matter is accreted and the angular momentum transferred. Combined with a comparatively weak magnetic field of $10^8-10^9$~G, rapid rotation is the hallmark of these "recycled" pulsars. Observations show that AMXPs and rotation-powered radio millisecond pulsars (MSPs) are evolutionarily connected. This link has been conclusively demonstrated for the so-called "transitional MSP" systems \citep[see, e.g.,][]{Archibald:2009aa, Bassa:2014uu, Archibald:2015vw, Papitto:2015wo}, which exhibit shifts from the rotation-powered radio MSP phase to the accretion-powered AMXP phase.

In the classic picture, the NS’s magnetic field truncates the accretion disc at some radius above the stellar surface. Matter is then funnelled along the field lines, impacting the star at (or near) its magnetic poles. The resulting accretion hotspots generate pulsed X-ray emission at the NS's spin frequency as the stellar spin rotates them in and out of view. 
Currently, about two dozen AMXPs have been discovered, covering a range of spin frequencies from a low of 105 Hz to nearly 600 Hz \citep[see][for extensive reviews]{Di-Salvo:2020va, Patruno:2021vs}. In most cases, outbursts last from a week to a few months, interspersed with long quiescent intervals. During an outburst, these systems often reach X-ray luminosities of $10^{36}\text{--}10^{37}$ erg s$^{-1}$, to which different emission regions contribute. A soft thermal component below a few keV is thought to arise from the inner area of the accretion disc or the NS surface. Meanwhile, higher energies feature a Comptonized spectrum, yielding a power-law–like shape that often exhibits a cutoff energy at tens of keV associated with the electron temperature of the Comptonized cloud or corona \citep[see, e.g.,][for a review]{Di-Salvo:2020va}.

A frequently observed phenomenon in AMXPs (and other NS LMXBs in particular) is thermonuclear, or type-I, X-ray bursts. These flashes occur when a critical column of accreted material ignites on the NS's surface, triggering a rapid thermonuclear runaway. These events are observed as short-lived (tens of seconds) bursts of X-ray emission. They can be extremely luminous, sometimes reaching the Eddington limit and driving a wind off the NS surface. Such bursts are referred to as photospheric radius expansion (PRE) bursts and can enable constraints on the source distance \citep[see][for more details on the topic]{Lewin:1993aa, Strohmayer:2006aa, Galloway:2017aa}. In the case of AMXPs, burst oscillations are usually observed to coincide with the star’s spin frequency \citep[see. e.g.,][]{Chakrabarty:2003aa}, offering insight into how the thermonuclear flame spreads and whether the burst emission is modulated by the same magnetic field geometry that channels accreting matter during the persistent state \citep[see][for a review]{Watts:2012aa}.

Coherent timing across single outbursts has revealed episodes of both spin-up \citep[due to accretion torque, see, e.g.,][]{Falanga:2005aa, Burderi:2006va, Burderi:2007tl, Papitto:2008ua, Riggio:2008wz, Riggio:2011vs} and spin-down \citep[likely from magnetically threaded disc regions exerting a braking torque or from electromagnetic losses during quiescence, see, e.g.,][]{Galloway:2002wz, Papitto:2007wp, Papitto:2011aa, Sanna:2020wv}. It is worth noting that timing noise can contaminate the measured spin frequency derivatives obtained from phase-connected timing, potentially masking genuine torque-driven spin variations and hindering accurate measurements in specific sources or during particular outbursts \citep[see, e.g.,][]{Patruno:2009vg, Hartman:2008uj}. When observed over different outbursts, AMXPs can be inspected for long-term spin and orbital evolution. Only nine AMXPs, including \igr{}, have been observed with high-time resolution instruments across different outbursts \citep[see, e.g.,][]{Di-Salvo:2020va}. Long-term spin-down evolution has been reported for five of these systems, with a spin-down frequency of the order of $10^{-15}$~Hz s$^{-1}$ \citep[see, e.g.,][and references therein]{Riggio:2011vs, Patruno:2017ah, Sanna:2018aa, Bult:2018ve, Illiano:2023aa}. The stability of the spin-down rate over the years suggests the loss of angular momentum via magnetic-dipole radiation as the most plausible explanation, which is expected for a rapidly rotating NS with a magnetic field. The range of measured spin-down rates is consistent with an average polar magnetic field between $10^8-10^9$~G, in agreement with
other estimates \citep[see, e.g.,][]{Mukherjee:2015td}. Moreover, for at least three of these systems there is evidence of the long-term evolution of the orbital period \citep[see][Riggio et al., in prep]{Sanna:2016ty, Bult:2020tu, Illiano:2023aa}, with SAX J1808.4$-$3658 showing the most puzzling behavior over the last seventeen years of monitoring that has captured nine outburst phases of the source \citep[see, e.g.,][for possible interpretations]{di-Salvo:2008uu, Burderi:2009td, Hartman:2009tq, Patruno:2012tw}.

\igr{} is a NS X-ray transient discovered by \inte{} in 2009 \citep{Baldovin:2009aa, Bozzo:2010aa}. Soon after its discovery, the detection of Doppler-modulated pulsations at approximately 245 Hz within a binary system with an orbital period of 3.47 hours allowed its classification as an AMXP \citep{Markwardt:2009aa}. The donor is a low mass star \citep{Papitto:2010aa}, consistent with the broader AMXP population, and transfers material that flows through an accretion disc before reaching the magnetically channelled region near the NS poles. Thermonuclear (type-I) X-ray bursts have been observed from \igr{} by different instruments \citep[see, e.g.,][]{Bozzo:2010aa, Papitto:2010aa, Papitto:2016wb}. None of the observed bursts showed evidence of photospheric radius expansion (PRE). This allows for setting only an upper limit to the distance of 6.9~kpc. Moreover, burst oscillations compatible with the source spin frequency were reported in a large fraction of the detected bursts \citep{Watts:2009aa, Altamirano:2010aa, Papitto:2016wb}.

The most accurate available source position is $RA=17h 51' 08.66''$, $DEC=-30^\circ57' 41.0\arcsec$ (1 sigma error of 0.6 arcsec), and was obtained with \chandra{} \citep{Nowak09}. Follow-up near-infrared observations identified a $K_s \sim 18$ candidate counterpart \citep{Torres:2009aa}, while only an upper limit of 0.1 mJy was reported in radio after scanning the same area \citep{Miller-Jones:2009aa}.

The average broadband X-ray spectrum during the outburst phase shows two main components, a soft blackbody-like component, likely originating from the inner disc or stellar surface, and a relatively hard component extending to tens of keV, which is well modelled by thermal Comptonisation in an electron cloud at temperatures ranging between 20-50 keV \citep{Papitto:2010aa, Bozzo:2010aa, Papitto:2016wb}. The detection of both broadened iron K$\alpha$ lines around 6–7 keV and a Compton hump at around 30 keV indicates that reflection processes occur close to the NS, suggesting an inner disc radius truncated around 40 km and an inclination between $38^\circ-68^\circ$ \citep[assuming a 1.4~M$_\odot$ NS; ][]{Papitto:2010aa}.

\igr{} was observed again in outburst for approximately one month starting March 23, 2015, \citep{Bozzo:2015aa}. The spectral and temporal properties of the source were remarkably similar to those of the 2009 outburst \citep{Papitto:2016wb}.

Renewed activity from \igr{} was reported by \inte{} around February 11, 2025 \citep{2025ATel17029....1S}, at a 28–60 keV flux of $1.4\times 10^{-10}$~erg~cm$^{−2}$~s$^{-1}$, suggesting the onset of the third known outburst of the source. A \nicer{} follow-up observation detected \igr{} at a 0.5-10~keV flux of $4\times 10^{-10}$~erg~cm$^{−2}$~s$^{-1}$ and recovered coherent X-ray pulsations at $\sim245$~Hz, confirming the new outburst of the source after a ten-year-long quiescence phase \citep{2025ATel17032....1N}. Here, we report on the spectral and temporal properties of \igr{} exploiting the 2025 datasets collected with \nicer{} and \nustar{}. Moreover, we reanalysed past archival \nustar{} and \xmm{} data to improve the constraints on the long-term evolution of the source.

\section{Observation and data reduction}

\subsection{\nicer{}}
\igr{} was observed by \nicer{} between February 11, 2025, 16:48 UTC and February 14, 2025, 13:06:29 UTC. We applied standard screening and filtering of the dataset using HEASoft version 6.34 and the \nicer{} Data
Analysis Software (NICERDAS) version 12 (2024-02-09 V012) using calibration version xti20240206. The screening obtained with the \textit{nicerl2} pipeline resulted in an exposure time of
approximately 9~ks distributed around twenty snapshots. 

The source plus background light curve (Fig.~\ref{fig:nicer_phase_fit}, top panel) shows a stable count rate during the monitoring, with an average count rate of 52 cts/s in the 0.3-10~keV energy range. No type-I X-ray burst events nor X-ray flaring episodes were recorded during the observation. We
corrected the \nicer{} photon arrival times for the motion of the Earth-spacecraft system with respect to the Solar System barycenter by using the FTOOLS \textit{barycorr} tool (DE-405 solar system
ephemeris) with the best available source position reported in Table 1 \citep{Nowak09}.

We extracted the spectral products using the \textit{nicerl3-spect} pipeline tool with the cleaned events in the 0.4-10~keV energy range. We applied the SCORPEON background model version 22 with the option \textsc{bkgformat=file}. To perform spectral fitting with enough statistics, we grouped the energy channels to have a minimum of 25 counts per energy bin (\textsc{GROUPSCALE = 25} and applied the option \textsc{GROUPTYPE=OPTMIN} \citep{Kaastra2016aa} from the FTOOLS \textit{ftgrouppha}.
We performed spectral fitting with XSPEC 12.14.0 \citep{Arnaud96}. 

\subsection{\nustar{}}

\nustar{} observed \igr{} (Obs.ID.
91101304002) on February 19, 2025, starting from 15:10 UTC
for a total coverage of $\sim$66.4~ks. After applying the standard screening
and filtering procedure with the data analysis
software \textsc{NUSTARDAS} from HEASOFT, version 6.33.2, we obtained a cleaned event data set with a total exposure of $\sim$37.6~ks. We extracted
source and background events from circular regions of radius 120'' and 140'' at the source position and in a source-free region of the same quadrant, respectively. We extracted light curves and spectra (including correlated response files) using the \textsc{nuproducts} pipeline. Furthermore, we used the \textit{lcmath} task to create background-subtracted light curves for each \nustar{} FPM and a cumulative one, resulting in a total mean count rate of $\sim12.4$~ cts/s in the 3-80 keV energy range.
As reported in Fig.~\ref{fig:nustar_phase_fit}, around $30$~ks from the beginning of the observation, \nustar{} reveals a sudden and intense increase in X-rays that resembles a typical type-I X-ray burst profile. For our analysis of the coherent X-ray pulsation, we removed data from a 200-s interval centred on the occurrence time of the X-ray burst. 
We applied barycentric corrections to the \nustar{} photon arrival times by using the \textsc{barycorr} tool (DE-405 solar system
ephemeris) with the best available source position reported in Table 1 \citep{Nowak09} and applying the latest clock correction file.

To be able to improve the ephemerides of \igr{} during its 2015 outburst, we analysed the archival \nustar{} observation (obs.ID. 90101001002) performed on April 8, 2015, starting from 20:36 UTC for a total coverage of $\sim$88.5~ks. Following the same procedure described earlier, we extracted
source and background events from circular regions of radius 120'' and 140'' at the source position and in a source-free region of the same quadrant, respectively. We applied the standard screening
and filtering procedure and extracted light curves and spectra (including correlated response files) for a total exposure time of $\sim$46.4~ks using the standard \nustar{} pipelines.
A type-I X-ray burst is observed close to the end of the observation. We excluded the event while performing the spin frequency timing analysis. Also for this dataset, we applied barycentric corrections using the \textsc{barycorr} tool with the source position reported in Table~\ref{tab:solution}.

\subsection{\xmm{}}
We extracted the \xmm{} observation of \igr{} performed on March 26, 2015, at 22:17 UTC for a total coverage of $\sim$76~ks (Obs.ID. 0770580301). 
Using the latest version of the \xmm{} scientific software  (SAS v.22.1.0), we extracted the events collected by the EPIC-pn camera operated in timing mode. Following \citet{Papitto:2016wb}, we first filtered the data from soft proton flaring episodes, then 
we extracted the source by selecting only events with the RAWX coordinate within the interval 27 to 47. We extracted the background filtering events within a three-pixel-wide region centred on the coordinate RAWX=4. The three type-I X-ray bursts observed during the \xmm{} observation were excluded while performing the spin frequency timing analysis. For the timing analysis, we reported the EPIC-pn photon arrival
times to the solar system barycentre by using the \textsc{BARYCEN} tool (DE-405 solar system ephemeris), adopting the best available source position reported in Table~\ref{tab:solution}.

\section{Results}

\subsection{Timing analysis}

To perform the timing analysis of the latest outburst of \igr{}, we started by correcting the \nicer{} and \nustar{} photon arrival times for the binary orbital motion under the hypothesis of a circular orbit \cite[see, e.g.,][]{Deeter:1981te, Burderi:2007tl}. As a starting point, we propagated the orbital ephemeris reported from the previous outburst \citep[see Table 3 in][]{Papitto:2016wb} and extrapolated the closest value of the time of passage of the ascending node ($T_{ASC}$) to begin looking for possible shifts in time. We then explored the orbital parameter space around the values reported by \citet{Papitto:2016wb}, varying $T_{ASC}$ by steps of 1~s within its 3$\sigma$ confidence level interval derived from the 2015 outburst. We searched for coherent pulsations for each set of orbital parameters by applying the epoch-folding method to both datasets. We divided the pulse profile into 20 phase bins and explored the frequency space around $v=244.83395112(3)$~Hz (the mean spin frequency during the 2015 outburst) with a step of $10^{-7}$~Hz for a total of 5001 steps. We recovered strong pulsed signals in both datasets, with the most prominent signal-to-noise ratio pulse profile corresponding to a common $T_{ASC}=60717.674078$~MJD and a frequency of $\nu=244.8339503$~Hz for both \nicer{} and \nustar{} datasets.

\begin{figure}
\centering
\includegraphics[width=0.45\textwidth]{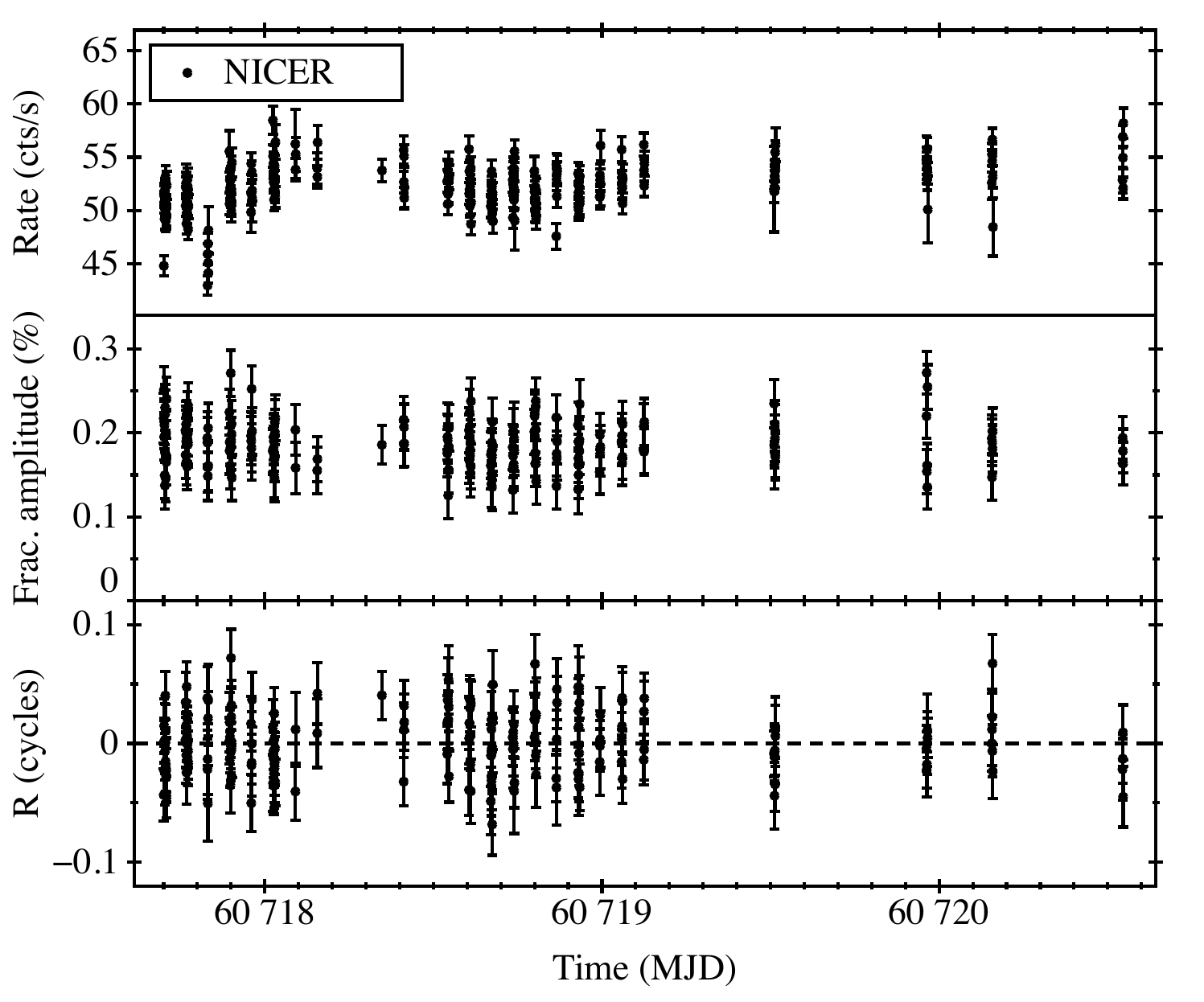}
\caption{\textit{Top panel -} \nicer{} 0.3–10~keV light curve of \igr{} during the latest outburst starting from February 11, 2025 (MJD 60717.7). The count rate is estimated by collecting 50 s of exposure time. 
\textit{Middle panel -} Temporal evolution of the fractional amplitude of the sinusoidal component used to model the source pulse profiles. \textit{Bottom panel -} Pulse phase residuals in units of phase cycles relative to the best-fitting solution.}
\label{fig:nicer_phase_fit}
\end{figure}

\begin{figure}
\centering
\includegraphics[width=0.45\textwidth]{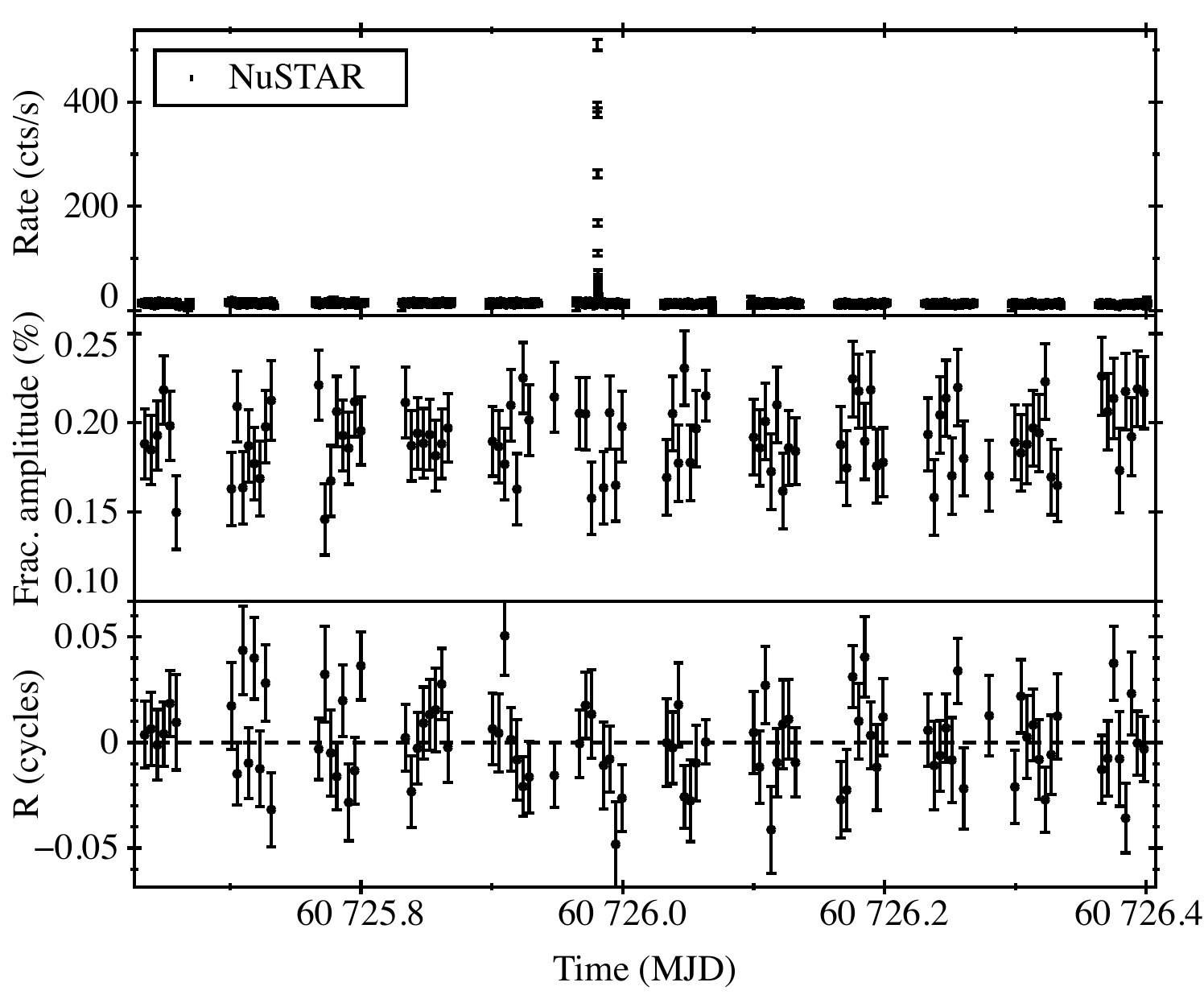
}
\caption{\textit{Top panel -} \nustar{} Background-subtracted light curve of the latest outburst of \igr{} as observed combining data collected by \nustar{} FPMA and FPMB (ObsID. 91101304002) starting from February 19, 2025 (MJD 60725.6). A type-I X-ray burst is detected at around MJD 60725.98.
\textit{Middle panel -} Temporal evolution of the fractional amplitude of the sinusoidal component used to model the source pulse profiles. \textit{Bottom panel -} Pulse phase
residuals in units of phase cycles relative to the best-fitting solution.}
\label{fig:nustar_phase_fit}
\end{figure}

Using the preliminary updated value of $T_{ASC}$ and the first guess of the latest spin frequency of the source, we demodulated the photon arrival times for the binary orbit. We created pulse profiles by epoch-folding 50 and 500-s long data segments, for \nicer{} and \nustar{}, respectively, considering the different count rates of the data sets. Using the corresponding mean spin frequency values obtained from the epoch-folding search method, we generated pulse profiles by considering 10 phase bins per interval to collect enough photons per bin within the selected time interval. 
We characterised each pulse profile with a sinusoidal function, from which we extracted the corresponding amplitude and fractional part of the phase residual. When required, we added a second harmonically related sinusoidal component to improve the modelling of the pulse profile. We retained only profiles with statistical significance greater than $3\sigma$.
We studied the temporal evolution of the derived pulse phases applying standard phase-coherent timing techniques \citep[see, e.g.,][and references therein]{Burderi:2007tl, Sanna:2016ty}. We repeated the process until no significant differential corrections were found for the model's parameters. In Table~\ref{tab:solution}, we reported the best-fitting parameters from the phase-coherent timing analysis. Following \citet{Papitto:2016wb}, we estimated the uncertainties on the spin frequency by adding the systematic uncertainty arising from the positional uncertainty in quadrature. The middle and bottom panels of Fig.~\ref{fig:nicer_phase_fit} and Fig.~\ref{fig:nustar_phase_fit} show the fractional amplitude and the best-fitting residuals for each time interval explored for \nicer{} and \nustar{}, respectively. 
The reduced $\chi^2_\mathrm{red}$ obtained from the joint fit and the pulse phase best-fit residuals distribution are compatible with the presence of timing noise, largely observed in several AMXPs \citep[see, e.g.,][and references therein]{Burderi:2006va, Papitto:2007wp, Riggio:2008wz, Riggio:2011vs, Patruno:2021vs}. A dedicated consistency check on the phase–residuals (power-spectrum and autocorrelation diagnostics) shows no statistically significant low-frequency (“red”) component, indicating that the excess variance is consistent with white noise in the pulse phases. To account for that, the uncertainties on the parameters reported in Table~\ref{tab:solution} are rescaled by a factor $\sqrt{\chi^2_\mathrm{red}}$ for a more realistic estimation \citep[see, e.g.,][]{Finger:1999vb}\footnote{We stress that the $\sqrt{\chi^2_\mathrm{red}}$ rescaling is a pragmatic convention—appropriate only when the unmodelled variance is temporally white—not a formally derived likelihood treatment.}.

\begin{figure}
\centering
\includegraphics[width=0.45\textwidth]{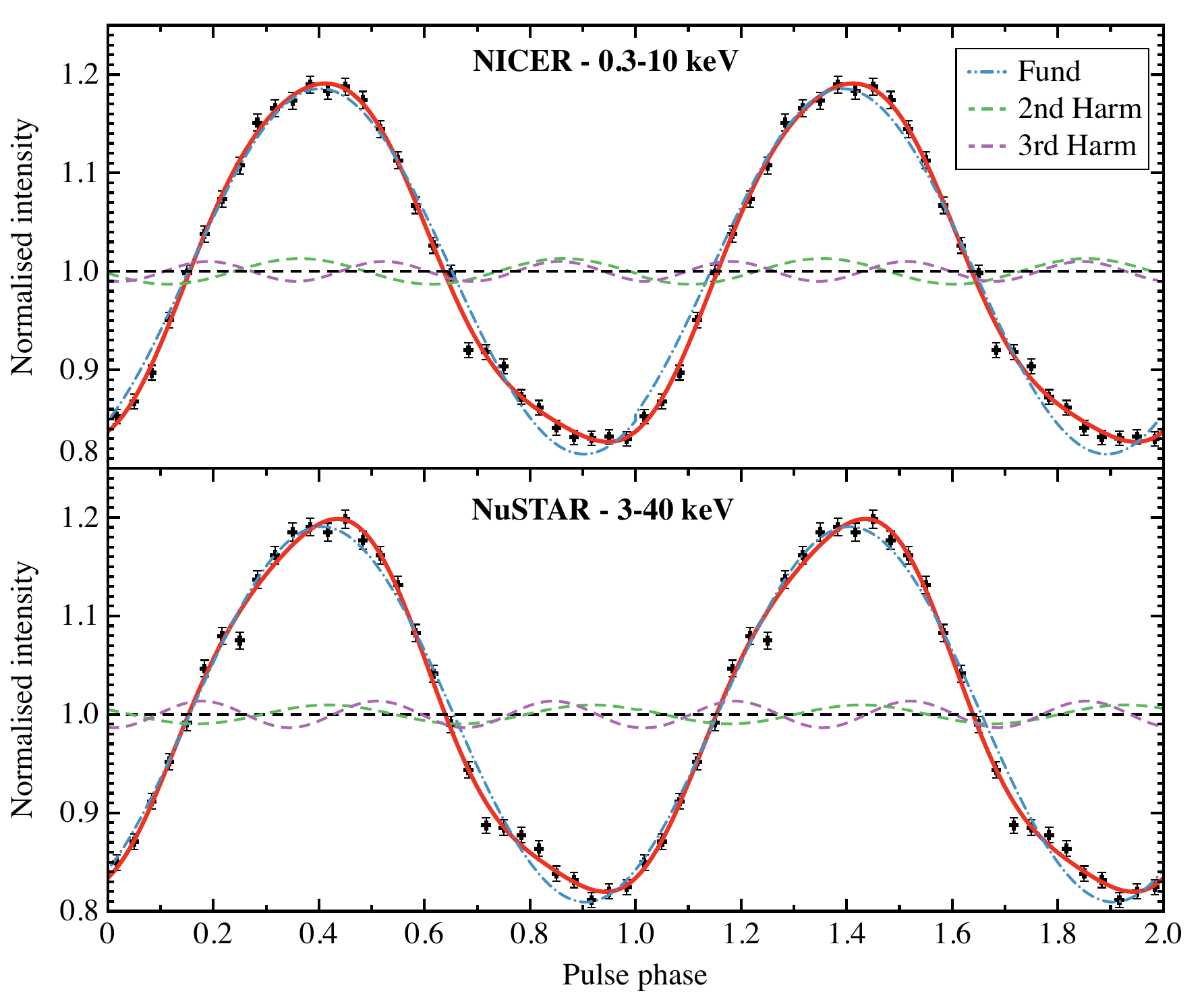}
\caption{\nicer{} 0.3-10~keV average pulse profiles (\textit{top panel}) and 3-40~keV average pulse profiles (bottom panel) of \igr{} generated by folding at the best-fit timing solution reported in Table~\ref{tab:solution}.
For both profiles, the best-fitting models (red solid lines) are the superposition of up to three harmonically related sinusoidal components represented in light blue, green, and purple from smaller to higher order, respectively. Two cycles of the pulse profile are shown for clarity.}
\label{fig:average_prof}
\end{figure}

Using the updated ephemeris reported in Table~\ref{tab:solution}, we created an average pulse profile with the highest statistical significance in the 0.3-10.0~keV and 3.0-40.0~keV energy bands for \nicer{} and \nustar{}, respectively. The profiles reported in Fig.~\ref{fig:average_prof} are well described by combining three harmonically related sinusoids with a dominant fundamental component. 

The \nicer{} profile (Fig.~\ref{fig:average_prof}, top panel) shows a background-corrected fractional amplitude of 19\%, 1.3\%, and 1.1\% for the fundamental, second, and third harmonics, respectively. Similarly, the \nustar{} profile 
is characterised by a background-corrected fractional amplitude of 19.5\%, 1.0\%, and 1.3\% for the fundamental, second, and third harmonics, respectively.

\begin{figure*}
\centering
\includegraphics[width=0.85\textwidth]{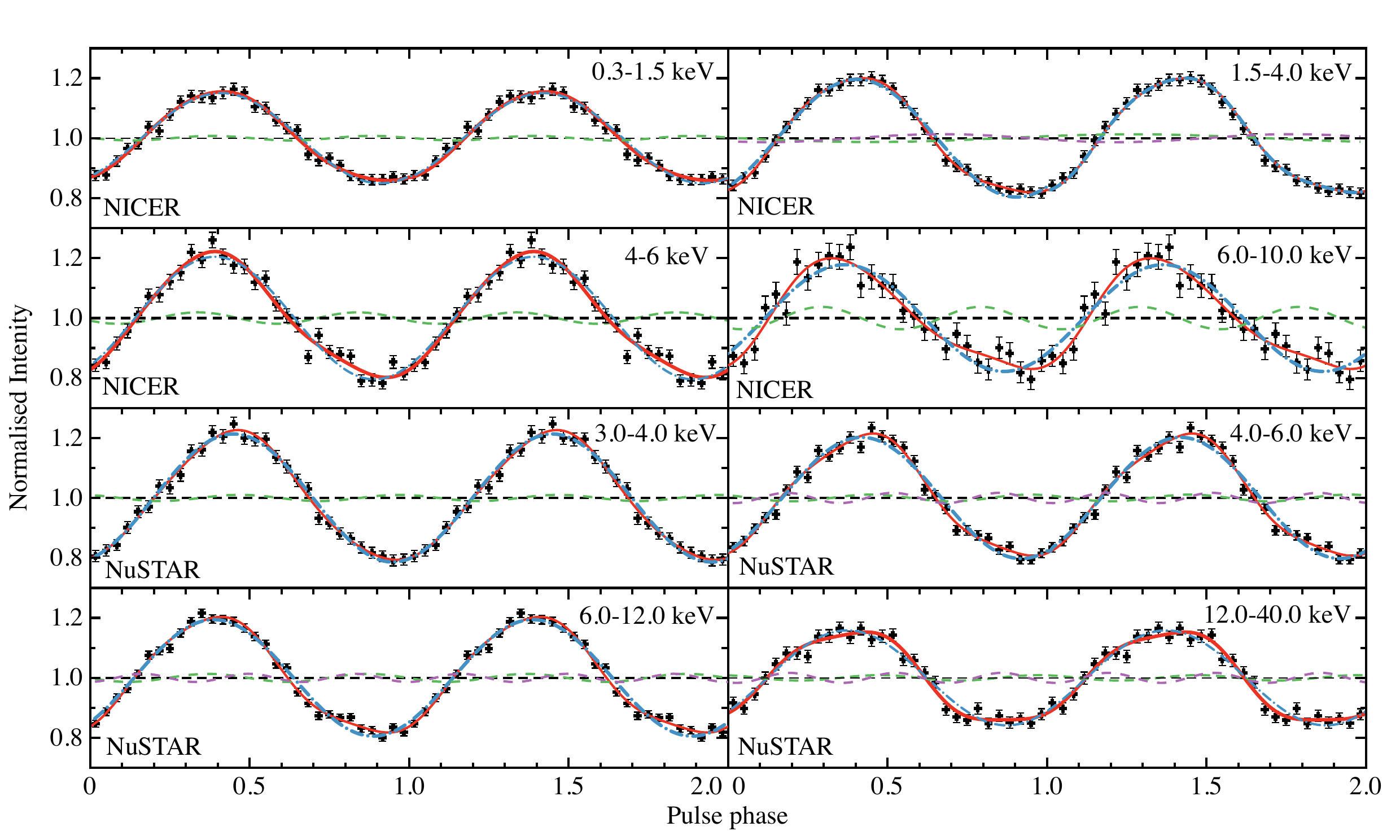}
\caption{Evolution of the \nicer{} and \nustar{} pulse profiles of \igr{} as a function of energy. Pulse profiles are generated by folding at different energy ranges after correcting the photon time-of-arrival for the most updated binary ephemerides reported in Table~\ref{tab:solution}. The best-fitting models (red solid lines) are the superposition of up to two harmonically related sinusoidal components. Two cycles of the pulse profile are shown for clarity. Colour coding follows the convention described in Fig.~\ref{fig:average_prof}.}
\label{fig:energy_profiles}
\end{figure*}

We investigated the energy dependence of the pulse profile for the two datasets. We selected four energy bands for \nicer{} (0.3-1.5~keV, 1.5-4.0~keV, 4.0-6.0~keV and 6.0-10.0~keV) and \nustar{} (3.0-4.0~keV, 4.0-6.0~keV, 6.0-12.0~keV and 12.0-40.0~keV) and folded the corresponding events using the parameters reported in Table~\ref{tab:solution}. The \nicer{} pulse profiles (top four panels in Fig.~\ref{fig:energy_profiles}) are well described as the superposition of two harmonically related sinusoidal components, except for the 1.5-4.0~keV where a third harmonic is required to improve the modelling of the profile. The background-corrected fractional amplitude of the fundamental component increases from $\sim16$\% at the lowest energies, and reaches a maximum of $\sim21$\%. The second harmonic shows a linear trend with energy, rising from $~1.2$\% to $\sim4$\%. Interestingly, only for the 1.5-4.0~keV interval, a third harmonic with a fractional amplitude of $\sim1.3$\% (similar to the second harmonic amplitude) is statistically required to improve the pulse profile fit.  

The \nustar{} pulse profiles (bottom four panels in Fig.~\ref{fig:energy_profiles}) are well described as the superposition of three harmonically related sinusoidal components, except for the 3.0-4.0~keV where the third harmonic is not statistically significant. The fractional amplitude of the fundamental component peaks at $\sim21.5$\% in the range 3.0-6.0~keV and decreases to $\sim16$\% at the highest energies. The amplitude of the second harmonic remains constant around $\sim1.1$\% as the energy increases. The profiles extracted at energies above 4.0~keV show the presence of a third harmonic with an amplitude of $\sim1.5$\%, stronger than the second harmonic.

To investigate the long-term evolution of the spin frequency and the orbital parameters of \igr{}, we improved the accuracy of the source ephemerides reported by \citet{Papitto:2016wb} (based on \xmm{}) for the 2015 outburst by including a \nustar{} observation performed during the same outburst but never reported so far. We performed the timing analysis following the procedure extensively described above. Results from the study are reported in Table~\ref{tab:solution} (third column). 

Finally, using the same \rxte{} dataset, we verified the accuracy of the ephemerides reported by \citet{Riggio:2011vs} by performing a standard phase-coherent timing analysis. The results shown in Table~\ref{tab:solution} (second column) are compatible within uncertainties with those reported by \citet{Riggio:2011vs}.

\begin{table*}
\caption{\label{tab:solution}Timing solutions for \igr{} during the observed outbursts.}
\centering
\begin{tabular}{l c c c}
\hline
 & 2009 & 2015 & 2025 \\
\hline
Parameters  & \rxte{} & \xmm{}+\nustar{} & \nicer{}+\nustar{}\\
\hline
\hline
R.A. (J2000) & \multicolumn{3}{c}{$17^h51^m08.66^s \pm 0.6^s$}\\
DEC. (J2000) & \multicolumn{3}{c}{$-30^\circ57' 41.0\arcsec \pm 0.6\arcsec$}\\
$P_\mathrm{orb}$ (s) &  12487.5118(4) & 12487.507(1) & 12487.505(2)\\
$x$ (lt-s) &0.275196(2) & 0.275192(4) & 0.275197(8)\\
$T_{ASC}$ (MJD/TDB) & 55088.0320286(6) &  57107.858808(1) & 60717.674090(3)\\
Eccentricity & $< 4 \times 10^{-5}$ (3$\sigma$~c.l.) & $ < 9 \times 10^{-5}$ (3$\sigma$~c.l.) & $ < 2 \times 10^{-4}$ (3$\sigma$~c.l.)\\
$T_0$ (MJD/TDB) & 55087.8 & 57107.9 & 60717.7\\
\hline
\hline
$\nu_0$ (Hz) & 244.83395161(3) & 244.83395112(3)& 244.83395034(3)\\
\hline
$\chi^2_\mathrm{red}$/dof & 1.48/7005 & 1.51/1541 & 1.19/266\\
\hline
\end{tabular}
\tablefoot{Orbital parameters and spin frequency evolution of \igr{} obtained from the analysis of the \nicer{} and \nustar{} observations from its latest outburst. $T_0$ represents the reference epoch for this timing solution. Uncertainties reported on the last digit correspond to the 1$\sigma$ confidence level. Uncertainties are also scaled by a factor $\sqrt{\chi^2_\mathrm{red}}$. We added a systematic uncertainty of $3\times 10^{-8}$~Hz in quadrature to the spin parameters to account for positional uncertainties.}
\end{table*}

\subsection{Spectral analysis}
This section presents the spectral analysis of \nicer{} and \nustar{} data collected during different phases of the 2025 outburst. Additionally, we examined the time-resolved spectroscopy of the thermonuclear (type-I) X-ray burst observed with \nustar{}.

\subsubsection{\nicer{} spectroscopy}

A time-averaged spectral analysis was performed using \nicer{} observations of \igr{} from February 11- 14, 2025. This approach was chosen as the source intensity remained nearly constant, showing only a gradual increase during these observations (top panel of Fig.~\ref{fig:nicer_phase_fit}). The 0.4–10 keV \nicer{} spectrum was initially modelled with a simple absorbed power-law with a blackbody component. The photoelectric absorption from the interstellar medium was described with the {\tt Tbabs} component in {\tt XSPEC}, with {\tt wilm} abundances \citep{Wilms00} and {\tt verner} cross-sections \citep{Verner96}.  This model provided a good fit, yielding a reduced chi-square ($\chi^2_\mathrm{red}$) of 1.05 for 954 degrees of freedom (hereafter dof). Alternatively, we found that a semi-phenomenological Comptonisation model {\tt ThComp} \citep{Zdziarski:2020aa} with a seed blackbody component can provide an equally good fit, resulting in a $\chi^2_\nu$ of 1.03 for 954 degrees of freedom. This model, expressed as {\tt Tbabs*(Thcomp*bbodyrad)} in {\tt XSPEC}, was adopted for further analysis of the NICER spectrum of \igr{}.

\begin{figure}
\centering
\includegraphics[width=0.42\textwidth]{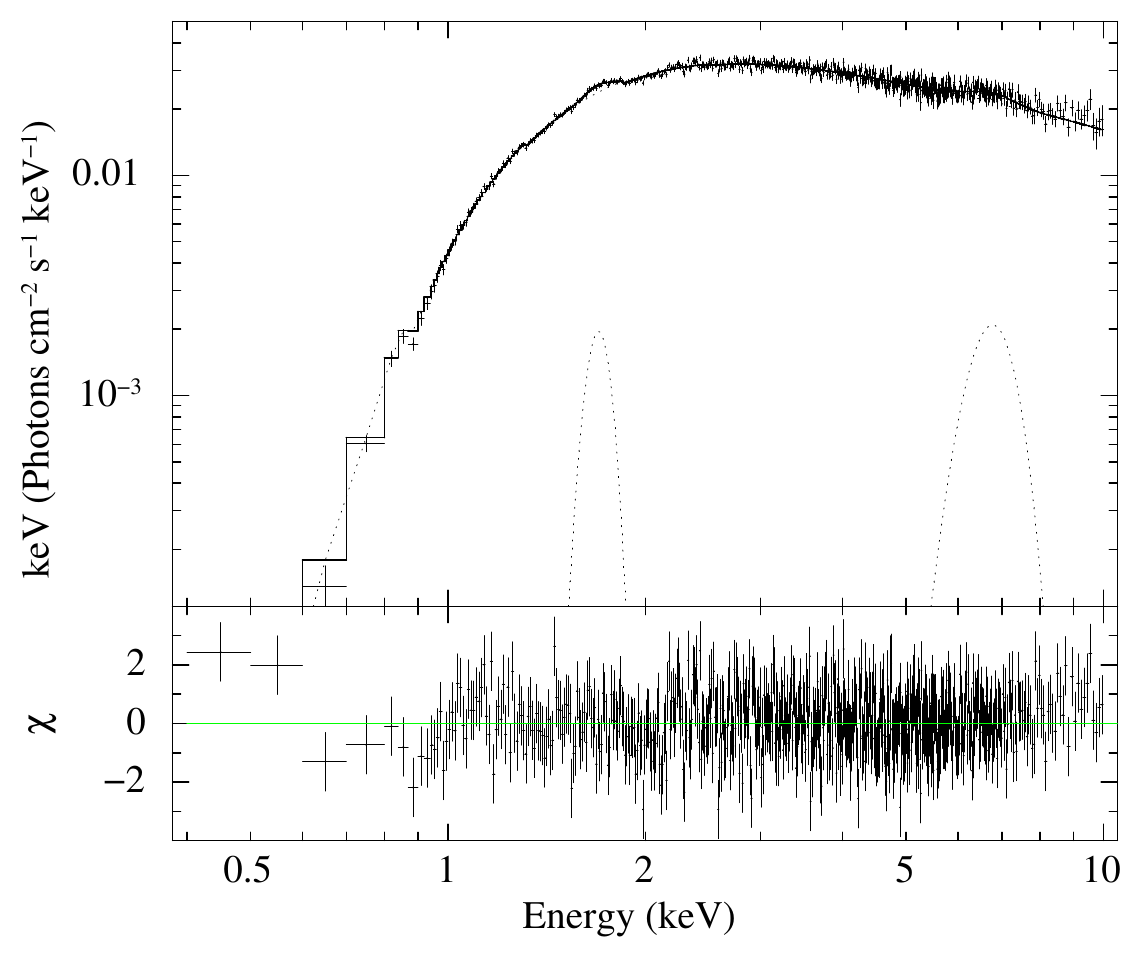}
\caption{The 0.4-10 keV \nicer{} spectrum fitted with an absorbed blackbody with a Comptonization model with two Gaussian emission line components (top panel). The corresponding spectral residuals are shown in the bottom panel.}
\label{fig:nicer-spec}
\end{figure}

Regardless of the modelling approach, emission residuals were observed around 1.7 keV. Adding a Gaussian emission component at 1.7 keV to the Comptonisation-based continuum model successfully accounted for the residuals, with an improvement in the $\chi^2$ ($\Delta\chi^2$) of 92.5 for three additional dof (corresponding to a significance $>9\sigma$). The best-fitting line centroid energy, $E_0$, width, $\sigma_E$, and  normalization, $N$, are $1.69 \pm0.02$ keV, $70\pm20$ eV, and $3.3\pm0.7 \times 10^{-4}$ cm$^{-2}$ s$^{-1}$, respectively. The equivalent width is 15$\pm$4~eV. Weak emission residuals were also observed in the 6--7 keV band. Including a second Gaussian line to account for these residuals improved the fit ($\Delta\chi^2$ = 28 for three additional dof, corresponding to a significance $>4\sigma$). The parameters for the second emission line are, $E_0 = 6.7\pm0.2$ keV, $\sigma_E = 0.6\pm0.2 $ keV, and $N=(4.3\pm 1.1) \times10^{-4}$ cm$^{-2}$ s$^{-1}$, with an equivalent width of $131\pm86$ eV. To assess the statistical significance of both emission features, we performed Monte Carlo simulations using {\tt XSPEC}'s {\tt simftest} command. For the 1.7 keV line, only 2 out of 1000 simulated spectra based on the null model showed a fit improvement equal to or greater than that observed in the real data, corresponding to a global p-value of 0.002, suggesting a line significance of 2.9$\sigma$. Moreover, for the 6.6 keV line, none of the 1000 simulations matched the observed improvement, implying a global p-value < 0.001 (significance $\ge$3.3$\sigma$). The model parameters from the best fit are also presented in Table~\ref{tab:spec-nicer}. Fig.~\ref{fig:nicer-spec} shows the NICER energy spectrum and corresponding residuals with the best-fitting model. 

Our study primarily focuses on time-averaged spectroscopy, however, we also analysed four individual NICER observation IDs from the 2025 outburst to evaluate potential variations in spectral parameters. During these observations, spanning February 11–14, 2025, the source unabsorbed flux gradually increased from $3.9 \times 10^{-10}$ to $4.2 \times 10^{-10}$ erg~s$^{-1}$~cm$^{-2}$ with an average flux of
$(4.03\pm0.01) \times 10^{-10}$ erg~s$^{-1}$~cm$^{-2}$ in the 0.5-10~keV as shown in Table~\ref{tab:spec-nicer}. The spectral parameters from each observation ID closely match the time-averaged values in the table.

\begin{figure}
\centering
\includegraphics[width=0.42\textwidth]{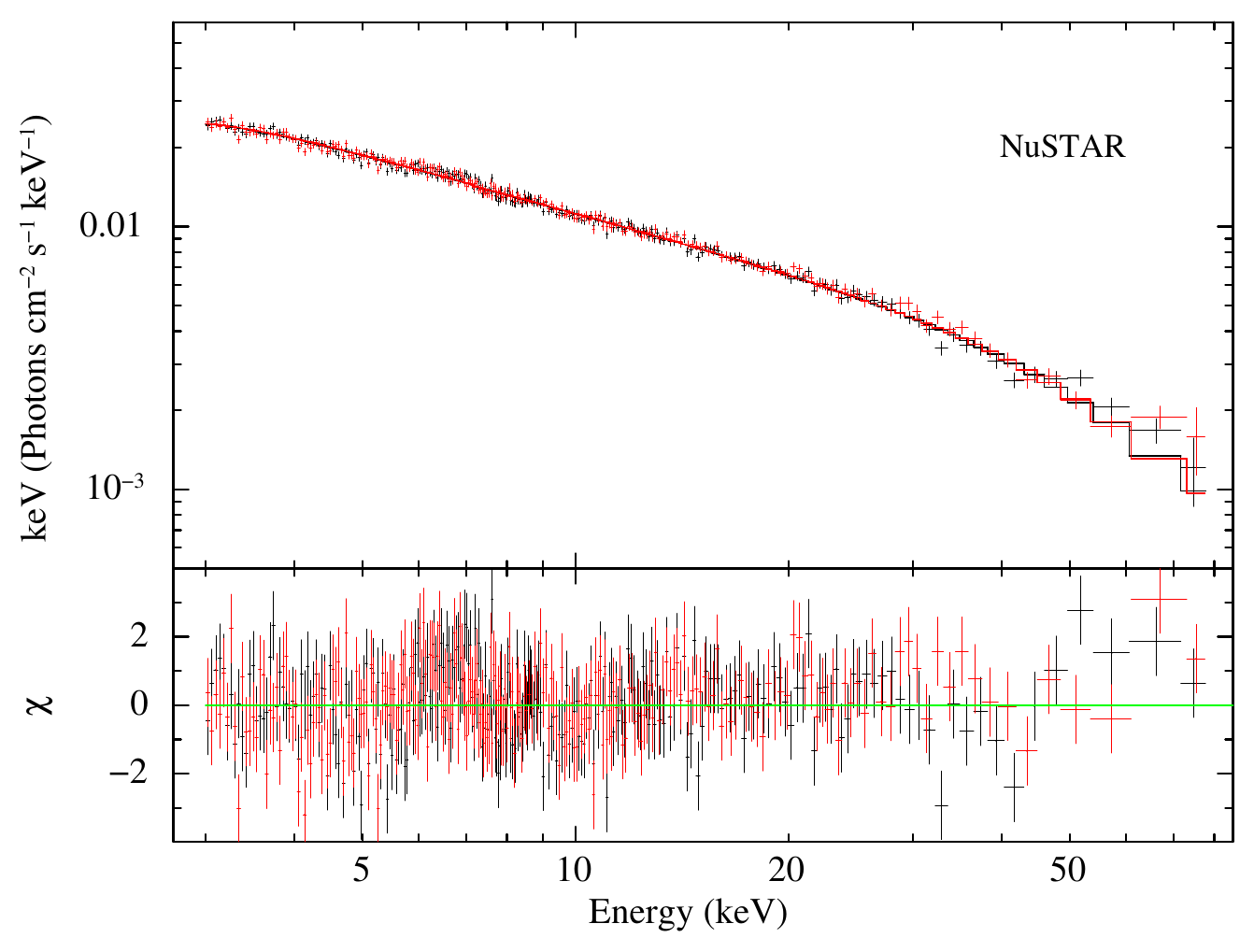}
\caption{Top panel shows the \nustar{} spectrum fitted with an absorbed blackbody with a Comptonization model. A weak hint of emission residuals in the 6--7~keV iron band is evident in the corresponding spectral residuals (bottom panel). }
\label{fig:nustar-spec}
\end{figure} 

\begin{table*}[]
\centering
\small	
\caption{\label{tab:spec-nicer}Best-fit spectral parameters for \igr{} during the 2025 outburst.}
\begin{tabular}{lccccc}
\hline
&   &\multicolumn{1}{c}{\nicer{}} & \multicolumn{2}{c}{{\nustar{}}}                       &\nicer{}+\nustar{}\\
Model         &Parameters                         &Model-I           &Model-I                &Model-II         &\\
\hline  
{\tt Tbabs}     &N$\rm{_H}$ (10$^{22}$ cm$^{-2}$)  &1.06$\pm$0.05     &1.06(fixed)      &1.06(fixed)          &1.08$\pm$0.05    \\    
{\tt bbodyrad}  &kT$\rm _{bb}$ (keV)               &0.52$\pm$0.04      &0.44$\pm$0.04         &                &0.49$\pm$0.01\\
                &Norm$\rm _{bb}$                   &290$^{+59}_{-75}$  &944$^{+1165}_{-367}$     &           &541$^{+65}_{-75}$\\
\\
{\tt ThComp}    &$\Gamma$                          &1.67$\pm$0.04        &1.79$\pm$0.01     &                 &1.62$\pm$0.01 $\&$ 1.79$\pm$0.01\\
                &kT$\rm {_e}$ (keV)                 &15.5$^{+12}_{-9}$   &17$\pm$1    &                      &17.4$\pm$0.8\\
\\
Emission Lines \\
Si              &Energy (keV)                           &1.69$\pm$0.02    &      &                           &1.69$\pm$0.02 \\
                &Width (keV)                            &0.07$\pm$0.02     &    &                             &0.07$\pm$0.02       \\ 
                &Norm (10$^{-4}$)                        &3.3$\pm$0.7     &     &                            &4.7$\pm$0.9       \\
                 
Fe              &Energy (keV)                           &6.7$\pm$0.2     &6.6$\pm$0.2             &            \\
                &Width (keV)                            &0.5$\pm$0.2     &0.6$\pm$0.2\\ 
                &Norm (10$^{-4}$)                        &4.3$\pm$1.1    &1.3$\pm$0.4 \\
\\
{\tt relxillCp}  &incl. (i) (deg)                        &     &                                       &$71_{-34}^{+9}$   \\
                 &R$_{\rm in}$ (R$_ {\rm g}$)                &     &                         &$89_{-49}^{+91}$            \\ 
                 &$\Gamma$                                    &     &                          &1.79$\pm$0.01      \\
                 &log$\xi$                                  &     &                             &$2.7_{-0.7}^{+0.6}$ \\
                 &log~N (cm$^{-3}$)                          &     &                            &$17.3_{-2.1}^{+2.3}$ \\
                 &kT${_e}$                                 &     &                            &17 (fixed) \\
                 &Afe                         &    &                                        &$4.2_{-3.2}^{+5.2}$ \\
                 &Ref$\rm_{frac}$                      &     &                            &$0.11_{-0.60}^{+0.12}$ \\

\\
{\tt Diskline}   &incl. (i) (deg)                        &     &                               &                &$>$32\\
                 &R$_{\rm in}$ (R$_ {\rm g}$)                &     &                         &           &$70_{-45}^{+53}$  \\ 
                 &Norm (10$^{-4}$)                          &     &                         &            &1.1$\pm$0.3 \\
                 
\\
                &Flux$^a$ (10$^{-10}$)       &4.03$\pm$0.01   &2.52$\pm$0.06    &2.70$\pm$0.08 \\
                &Flux$^b$ (10$^{-10}$)                  &   &6.67$\pm$0.03                           &6.75$\pm$0.05   &6.67$\pm$0.03 \\
                &$\chi^2_\mathrm{red}$ (dof)       &0.91 (948)    &0.99 (1479)                  &0.99(1478)            &0.97(2332)\\
\hline
\end{tabular}
\\
\tablefoot{Spectral parameters obtained by fitting the \nicer{} and \nustar{} observations. Model-I corresponds to an absorbed blackbody with a Comptonization component with Gaussian line component(s), whereas Model-II is an absorbed self-consistent reflection model ({\tt Tbabs*relxillCp}). Uncertainties are calculated at the 90\% confidence level.  
\tablefoottext{a}{Flux and} \tablefoottext{b} {Flux correspond to total model fluxes in the 0.5-10 and 0.5-100 keV ranges in units of erg~s$^{-1}$ cm$^{-2}$.}} \\
\end{table*}

\subsubsection{Semi-phenomenological and physical reflection modeling of \igr{} with \nustar{}}  
\label{sec:refl}
We extracted the burst-free spectrum from the FPMA and FPMB modules of \nustar{} using observations conducted on February 19–20, 2025. The 3–79 keV spectrum was modelled with an absorbed blackbody combined with a Comptonized component ({\tt Thcomp}), with the column density fixed at $1.06\times 10^{22}$ cm$^{-2}$, as estimated from the \nicer{} analysis. This model adequately represented the pulsar's energy continuum, with a $\chi^2_\nu$ of 1.02 for 1482 dof. A weak indication of emission residuals in the 6–7 keV range was also detected with \nustar{} (Fig.~\ref{fig:nustar-spec}). Incorporating a Gaussian emission line into the continuum model improved the fit with a $\Delta\chi^2$ of 48 for three additional dof. The best-fit parameters and their uncertainties at the 90\% confidence level are listed in Table \ref{tab:spec-nicer}. 

To explore the evolution of the outburst, we compared the unabsorbed flux in the 0.5--10 keV band obtained from \nicer{} and \nustar{}. During the \nustar{} observation, the flux had decreased by approximately 35 to 40\% compared to the \nicer{} measurements taken between February 11–14, 2025 (Table~\ref{tab:spec-nicer}). This decline indicates that the \nustar{} observation occurred during the decay phase of the 2025 outburst of \igr{}. Furthermore, variations in spectral parameters were observed between these observations (Table~\ref{tab:spec-nicer}). 

We further investigated the spectral behaviour during the decay phase of \igr{} with \nustar{}, using the self-consistent {\tt relxillCp} reflection model \citep{Garcia2014ApJ...782...76G, Dauser2014MNRAS.444L.100D}.  This physical model accounts for relativistic reflection from an incident {\tt nthcomp} continuum. Key model parameters include the photon index ($\Gamma$), the accretion disc's ionisation parameter (log$\xi$), disc density (log~N), inclination angle (i), inner and outer disc radii, and iron abundance relative to solar (Afe).  The disc emissivity is described by indices q1 and q2. For simplicity, we assumed a single emissivity profile, fixing q1 and q2 to the standard value of 3. In our fit, the outer disc radius was set to 1000 gravitational radii ($\rm R_g = GM/c^2$), with a break radius $\rm R_{br}$ at 400 $\rm R_g$. The dimensionless spin parameter was frozen at 0.7 \citep{Miller2011ApJ...731L...5M, Jaisawal2019ApJ...885...18J}. The choice of spin parameter, whether set to 0 or 0.7, does not significantly affect the derived spectral parameters. Based on the semi-phenomenological modelling with {\tt ThComp}, the electron temperature ($kT_e$) was fixed at 17~keV during the reflection modelling.  Other parameters, such as the inclination angle, inner disc radius, accretion disc density (log~N in cm$^{-3}$), ionisation parameter, relative iron abundance, reflection fraction parameter (Ref$\rm_{frac}$), and the model normalisation, were left free to vary. 

The 3-79 keV \nustar{} spectrum was self-consistently described using an absorbed {\tt rexillCp} model with a fixed absorption column density of 1.06$\times$10$^{22}$~cm$^{-2}$. This model also yielded a $\chi^2_\nu$ close to 1, indicating a good fit. Moreover, it also accounted for the weak emission residuals in the 6--7 keV range. The results suggest the presence of moderately ionised reflection, with log$\xi$ of $2.7_{-0.7}^{+0.6}$. The relative iron abundance is poorly constrained, with values ranging between 1 and 9.
The fit further indicated a high inclination angle of $71_{-34}^{+9}$ degrees, suggesting a lower limit of the inclination angle greater than $37^\circ$. The inner disc radius was constrained to a range of (40–180)$\rm R_g$, corresponding to (82-370)~km, assuming a canonical NS of 1.4 solar mass. The spectral parameters are given in Table~\ref{tab:spec-nicer}. Uncertainties on the parameters were estimated for a 90\% confidence interval using Markov Chain Monte Carlo ({\tt MCMC}) chain with the Goodman-Weare algorithm in {\tt XSPEC}. This simulation employed 2$\times$10$^6$ chain steps and 20 walkers, discarding the first 1$\times$10$^6$ steps as burn-in and transient phases. The {\tt MCMC} corner plot is shown in Fig.~\ref{fig:corner_plot}.

\subsubsection{Broadband spectroscopy with \nicer{} and \nustar}

Although \nicer{} and \nustar{} observations were not simultaneous, we jointly fitted their data better to constrain the inclination angle and inner disk parameters. We employed an absorbed Comptonized {\tt Thcomp} model combined with a blackbody component. Initially, this model did not adequately describe the spectra. To address potential spectral evolution between the two datasets, the photon index in the {\tt Thcomp} model was allowed to vary independently for NICER and NuSTAR. Additionally, a Gaussian feature was incorporated to represent a 1.7 keV emission line.
To further investigate the iron line feature, a {\tt Diskline} component was incorporated. The parameters of the {\tt Diskline}  include the emissivity index, which characterises the power-law dependence of the disk's illuminated emissivity profile, scaling as $r^{-\beta}$, along with the inner and outer disk radii and the disk inclination \citep{Fabian:1998aa}. For our analysis, the emissivity index was fixed at a value of 3. The smearing parameters derived from the {\tt Diskline}  model were consistent with those obtained using the self-consistent reflection model from the NuSTAR data using {\tt relxillCp} (see Table 2). The diskline model provided improved constraints, with the inner disk radius estimated in a range of (24-122)$\rm R_g$, representing a modest improvement compared to the values from the NuSTAR fit  (Table 2). The lower bound of the inclination angle was estimated to be 32 degrees without an upper constraint.
Other model parameters, such as the photon index derived from the {\tt Thcomp} model, were consistent with the values obtained from the individual fitting of NICER and NuSTAR. The blackbody emission parameters also match those from the individual fits.

\begin{figure}
\centering
\includegraphics[width=0.4\textwidth]{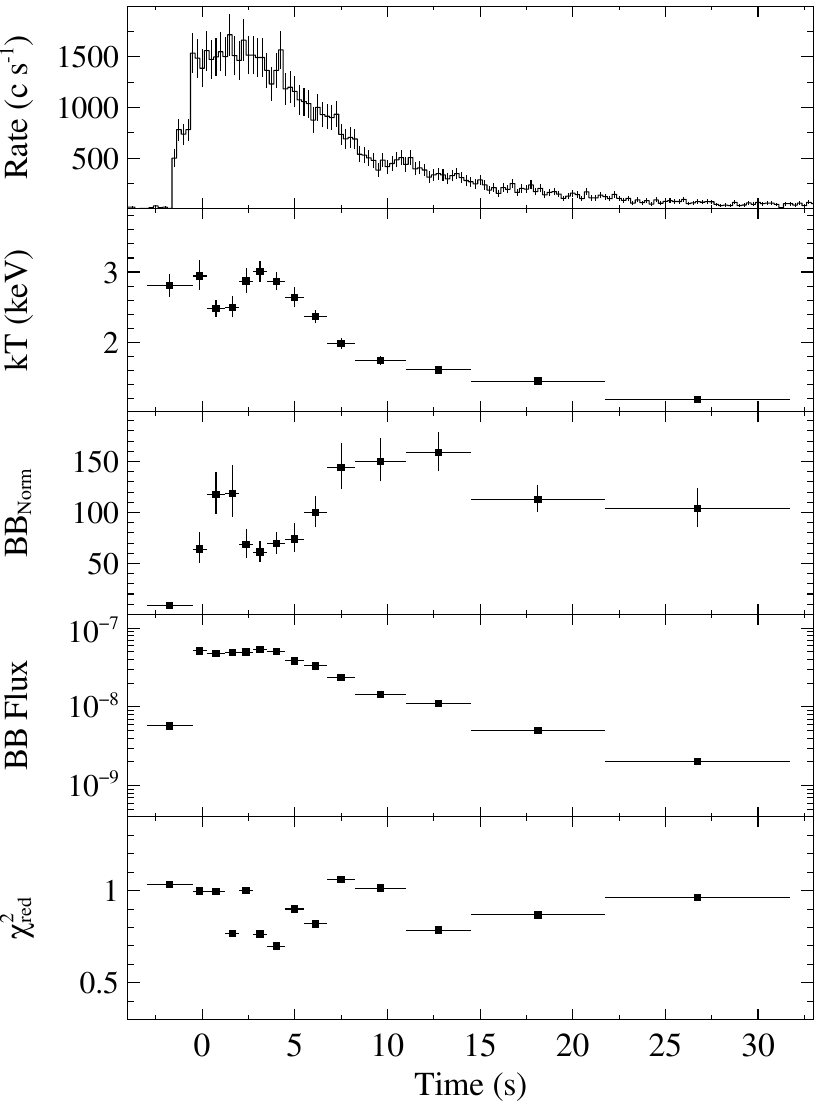}
\caption{Burst time-resolved spectral parameters. The top panel shows the burst light curve in the 3-79 keV range. The blackbody temperature (kT) and its normalisation (BB$\rm _{Norm}$) are shown in the second and third panels. The bolometric blackbody flux in the 0.1-100 keV range is presented in the fourth panel in the unit of erg~s$^{-1}$ cm$^{-2}$, whereas the fifth panel shows the value of $\chi^2_\mathrm{red}$ from the fit.}
\label{fig:burst-spec}
\end{figure} 

\subsubsection{Burst time-resolved spectroscopy}

A thermonuclear (type-I) X-ray burst was detected during the \nustar{} observation at an onset time of MJD 60725.98235. The burst reached an average peak count rate of $1500 \pm 200$ c~s$^{-1}$ in the 3–79 keV range light curve at a bin time of 0.25~s (see top panel of Fig. \ref{fig:burst-spec}). The profile exhibited a fast rise time of 2~s, followed by an exponential decay time of 8.5~s. The burst spanned a total duration of approximately 25~s, during which the count rate remained above 10\% of the peak burst rate. 
During the rise phase, the burst shows a pause-like structure, similar to those observed in SAX J1808.4-3658 \citep{Bult2019ApJ...885L...1B} and MAXI J1807+132 \citep{Albayati:2021aa} with NICER.

Before studying the burst time-resolved spectroscopy, we extracted a pre-burst spectrum in the same orbit of the burst with an exposure time of 1250~s. The pre-burst spectra from FPMA and FPMB detector modules are fitted using an absorbed blackbody model with a Comptonized {\tt ThComp} component.
The parameters obtained from the pre-burst fit are consistent with the values reported in Table~\ref{tab:spec-nicer} from burst-free persistent emission from \nustar{}. Our pre-burst parameters (at $90\%$ confidence level) are: blackbody temperature kT$\rm _{bb}$=0.6$\pm$0.1~keV, Norm$\rm _{bb}$=248$^{+404}_{-102}$, ThComp$_\Gamma$=1.84$\pm$0.01, ThComp$\rm_{kT{_e}}$=19.3$\pm$4.5, and the flux in 0.5-100 keV = (6.62$\pm$0.05)$\times$10$^{-10}$~erg~s$^{-1}$ cm$^{-2}$. The column density was fixed at 1.06$\times$10$^{22}$~cm$^{-2}$.

We performed time-resolved spectroscopy to investigate the nature of the observed X-ray burst with \nustar{}. We generated 14 time-resolved spectra to successively cover the burst's rise, peak, and decay phases, each containing a minimum of 1200 counts per interval. The exposure times for these intervals range between 0.75 and 10~s, with a median exposure of 1.125~s. Each burst time-resolved spectrum using data from FPMA and FPMB modules was fitted in the 3-30 keV range with an absorbed blackbody model at a fixed value of the above pre-burst emission parameters and absorption column density. This approach allowed us to incorporate the contribution of persistent accretion emission. The model successfully described the spectra of each time interval. Due to acceptable values of $\chi^2_\mathrm{red}$, we did not include a scaling factor to probe any changes in the pre-burst emission during the burst. The scaling factor is considered in the case of the variable persistent emission method, which accounts for the observed soft excess during X-ray bursts after the burst blackbody component (see, e.g., \citealt{Worpel2013, Jaisawal2024}).  As shown in Fig.~\ref{fig:burst-spec}, a high blackbody temperature of approximately 3~keV was detected around the rise and peak phases of the burst. This temperature exhibited an exponential decline during the decay phase. The blackbody normalisation increased during the rise and decay phases, reaching a maximum value of 160$\pm$20, corresponding to a blackbody emission radius of 8.7$\pm$0.6~km, assuming a distance of 6.9~kpc. The X-ray burst did not show any evidence of photospheric radius expansion. The burst peak flux in the 0.1--100 keV range is estimated to be (5.4$\pm$0.4)$\times$10$^{-8}$~erg~s$^{-1}$ cm$^{-2}$.  The observed flux matches well with the peak values detected in a range of (3--6.7)$\times$10$^{-8}$~erg~s$^{-1}$ cm$^{-2}$ from a sample of ten X-ray bursts with \rxte{} \citep{Altamirano:2010aa}. We also calculated the burst fluence by integrating the bolometric blackbody flux over each time interval from time-resolved spectroscopy. This resulted in a burst fluence of (5.1$\pm$0.1)$\times$10$^{-7}$~erg~cm$^{-2}$. The uncertainties on the burst parameters are given for a 68\% confidence level.   

\subsubsection{Burst oscillations}

\begin{figure}
\centering
\includegraphics[width=0.4\textwidth]{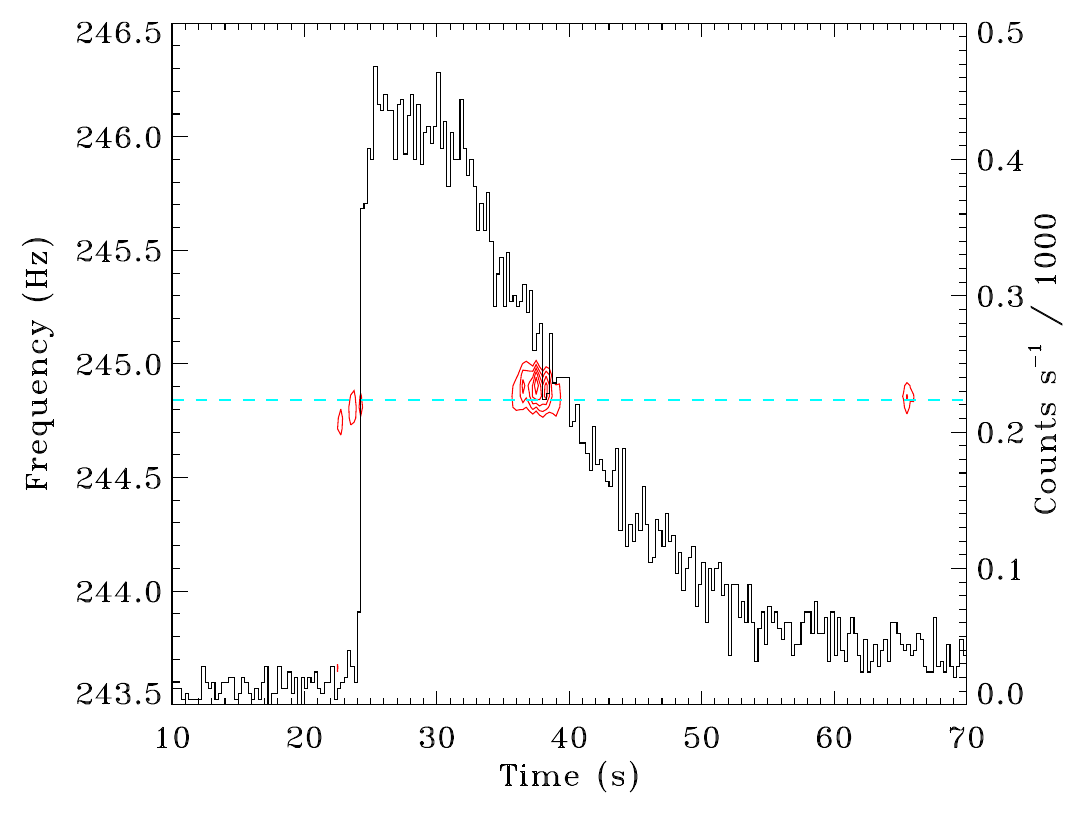}
\caption{Contours of constant $Z^2$ power are shown as a function of time and frequency from our burst oscillation search. Contours are plotted at $Z^2$ levels from 14 to 34 in steps of 4. The horizontal dashed line (cyan) denotes the measured spin frequency.}
\label{fig:burst-osc}
\end{figure}

\begin{figure}
\centering
\includegraphics[width=0.4\textwidth]{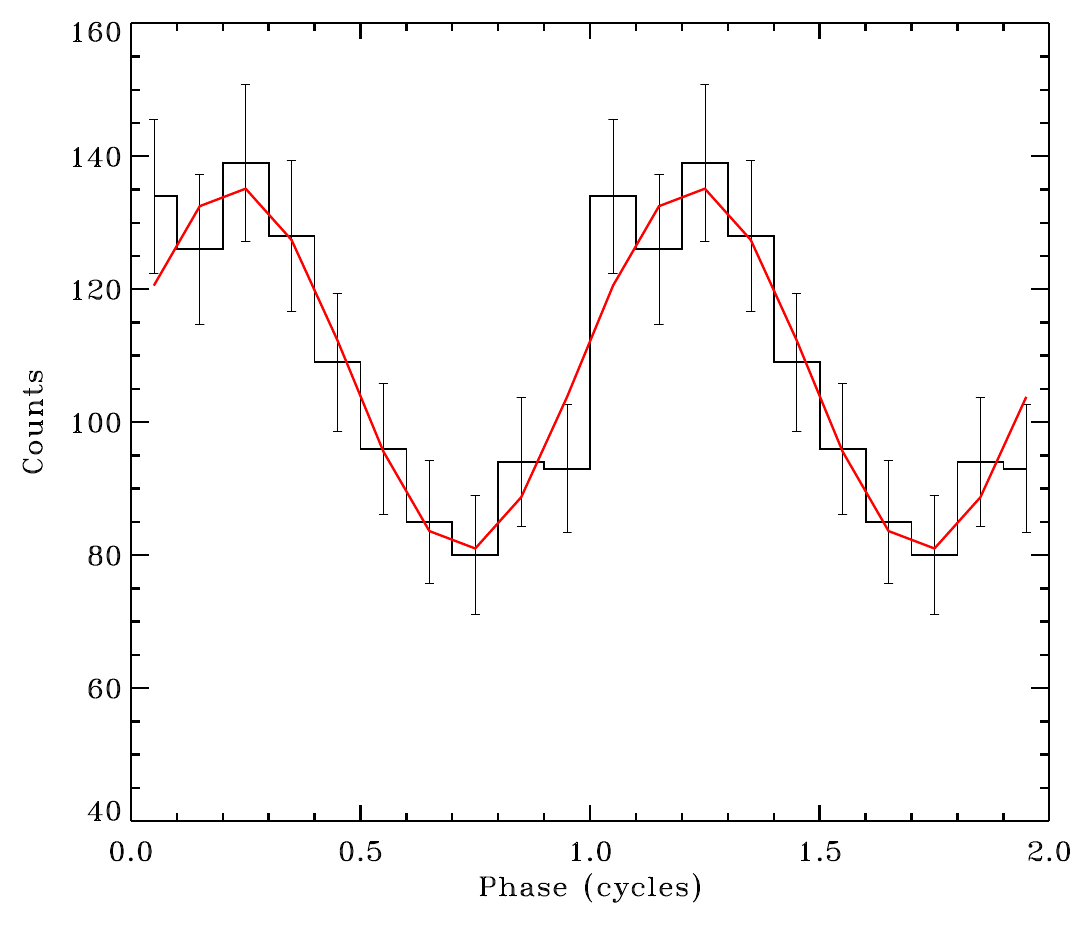}
\caption{Folded pulse profile from the 4 s interval which yielded the peak $Z^2$ power in our search. The best-fitting sinusoidal model is also plotted (red).}
\label{fig:burst-osc2}
\end{figure}

We searched the NuSTAR burst for oscillations since previously observed X-ray bursts from \igr{} have shown burst oscillations \citep{Altamirano:2010aa}. We used data in the 3-40 keV band and performed a sliding time window search. We selected a time range of $\sim$ 60 s around the burst and computed FFT power spectra using 4~s intervals.  The whole window begins approximately 14~s before the burst rise. For each 4~s interval, we computed the $Z^2$ statistic in a 3~Hz band of frequencies centred around the measured spin frequency. We slide the window by 1/4 of a second to obtain the next time window for the search. We found a strong candidate interval with a peak $Z^2$ power of 38.7 at a frequency consistent with the measured spin frequency.  This power value has a single trial chance probability of $4.0 \times 10^{-9}$.  We used a frequency resolution of 0.01 Hz for a total of 30 measurements across the searched frequency band. We also searched $226$ intervals, giving a total of $301\times226 = 67.424$ trials for the search, resulting in an estimate of the chance probability of $2.7 \times 10^{-4}$, which is a 3.5$\sigma$ detection. Fig.~\ref{fig:burst-osc} shows the contours of $Z^2$ power relative to the burst intensity profile. To check that this significance is not overstated by the strong overlap between adjacent windows and the correlations among closely–spaced frequencies, we ran 50.000 Monte-Carlo bursts that reproduce the observed envelope and were processed through the identical search pipeline.  Only two simulations reached a peak power equal to or larger than the real one, implying a global probability of $p_{\rm MC}=4.0\times10^{-5}$ (equivalent to $3.9~\sigma$), fully consistent with a highly significant detection once the true number of independent trials is taken into account. 
The position of the interval showing the highest $Z^2$ power is consistent with previous measurements of bursts with RXTE \citep[see][Fig. 5]{Altamirano:2010aa}.  We folded the peak $Z^2$ interval from our search using the measured frequency and fit a sinusoid to the resulting pulse profile in order to measure the pulsed amplitude.  A sinusoid fits the profile well, and the resulting amplitude is $25\pm 5 \%$. The folded profile, along with the best-fitting model, is shown in Fig.~\ref{fig:burst-osc2}.

\section{Discussion and Conclusions}

We presented the temporal and spectral properties of the
AMXP pulsar \igr{} during its 2025 outburst obtained by investigating the available \nicer{} and \nustar{} datasets. 

\subsection{Pulse profile energy dependence}

The average pulse profiles, which collect all available events in the 0.3-10~keV and 3.0-40.0~keV bands for \nicer{} and \nustar{}, respectively, are shown in Fig.~\ref{fig:average_prof}. Both best-fit models required three harmonics to successfully describe the pulse profiles, with a dominant fundamental component characterised by a fractional amplitude of $\sim19$\% in the two datasets. The second and third harmonics, both showing a similar fractional amplitude of $\sim1$\%, are required to model small asymmetries of the profiles. This is consistent with the modelling of the pulse profiles of the previous outburst, where, due to higher statistics, a fourth harmonic was significantly detected \citep[see, e.g.,][]{Papitto:2010aa, Riggio:2011vs, Papitto:2016wb}. Harmonically rich pulse profiles are not standard among AMXPs, so far, apart from \igr{}, only XTE~J1807$-$294 \citep[see, e.g.,][]{Patruno:2010wi}, Swift J1749.4$-$2807 \citep[][]{Altamirano:2011uq, Sanna:2022tt}, MAXI J1957+032 \citep{Sanna:2022vi}, IGR J16597$-$3704 \citep{Sanna:2018td}, and IGR J17591$-$2342 \citep{Sanna:2018wh, Kuiper:2020tl} required three or more components. 

Fig.~\ref{fig:energy_profiles} shows the energy dependence of the pulse profiles obtained by folding \nicer{} and \nustar{} datasets with the ephemerides reported in Table~\ref{tab:solution} in different energy bands. The profile amplitude, dominated by the fundamental component,  increases with energy with a common maximum of $\sim21$\% for the two datasets in the band 3.0-6.0~keV. This is compatible with what was reported from the previous outbursts from the \rxte{} and \xmm{} observations of the source. The observed fractional amplitude is among the highest observed so far in AMXPs, other systems showing large fractional amplitude (above $15\%$) are IGR J17379$-$3747 \citep[see, e.g.,][]{Sanna:2018tx, Bult:2019tr} SWIFT J1749.4$−$2807 \citep[see, e.g.,][]{Altamirano:2011uq, Sanna:2022tt}, XTE J1807$-$294 \citep{Patruno:2010wi}, and IGR J17591$-$2342 \citep{Sanna:2020wv}. According to \citet{Poutanen:2006aa}, large fractional amplitudes could reflect an inclination angle $> 45^\circ$ for rotators with relatively small misalignment angles ($< 20^\circ$). It is worth noting that for a two-hot-spot scenario, the maximum pulse amplitude is predicted to occur when the sum of the two angles is close to $90^\circ$. For \igr{}, this is consistent with the constraint on the inclination ($38^\circ-68^\circ$) reported by \citet{Papitto:2010aa}, as well as from the results discussed in Sec.~\ref{sec:refl}, from the modelling of the reflection continuum spectrum.

Interestingly, the harmonic content of the profile increases with energy. Both the second and the third (when detected) harmonics increase their amplitude with energy. The 6.0-10.0~keV \nicer{} pulse profile appears significantly affected by the increase in the second harmonic component. The third harmonic is detected in a few, relatively high-energy intervals, where its amplitude is similar to that of the second harmonic. This can be an observational bias related to the number of photons in the selected intervals. As shown in \citet{Papitto:2016wb} and \citep{Riggio:2011vs}, increasing the statistics allows us to significantly detect the third (and even a fourth) component at similar energy ranges, where the component shows a fractional amplitude around 1\% or less.

The energy dependence of the fundamental pulse amplitude of \igr{} resembles that of other AMXPs such as SAX J1748.9$-$2021 \citep{Patruno:2009tr, Sanna:2016ty}, 
IGR J00291+5934 \citep{Falanga:2005aa, Sanna:2017tx}, XTE J1807$-$294 \citep{Kirsch:2004tg, Patruno:2010wi}, XTE J1751$-$305 \citep{Falanga:2007ub}, and IGR J17494$-$3030 \citep{Ng:2021aa}. This behaviour may be ascribable to strong Comptonization of the beamed radiation, required to extend the pulsed spectrum to higher energies \citep[see, e.g., ][and references therein]{Falanga:2007ub}.

\subsection{Long-term spin evolution}
\label{sec:spin}

We performed phase-coherent timing analysis of the previous two outbursts of \igr{} detected since its discovery, obtaining consistent results with previous reported analysis \citep[see, e.g.,][]{Papitto:2010aa, Riggio:2011vs, Papitto:2016wb}. To determine possible secular evolution of the spin frequency, we combined previous outbursts with the updated ephemeris of the source obtained with \nicer{} and \nustar{} (Table~\ref{tab:solution}). We adopted a linear model to describe the temporal evolution of the three spin frequency values, obtaining a best-fit reduced $\chi^2_\mathrm{red}\sim 154$ with one dof and a spin frequency derivative of $\dot{\nu}_{sd}=-2.3(1.1)\times 10^{-15}$~Hz/s. Although the general trend shows a clear decrease in the spin frequency, the large value of reduced chi-squared indicates that a simple linear spin-down model is not statistically acceptable. This discrepancy most likely arises because the spin frequency evolves due to the transient accretion-driven torques associated with the active accretion phases during the outbursts, which a constant linear spin-down assumption cannot adequately capture. We further investigated the frequency evolution by accounting for the spin-up phases during the 2009 and 2015 outbursts. We estimated independently the frequency variation $\Delta \nu_{2009-2015}$ and $\Delta \nu_{2015-2025}$, comparing the end of the first outburst with the beginning of the second, and the end of the second with the beginning of the third outburst, respectively. Following \citet{Riggio:2011vs}, for the 2009 outburst we assumed a duration of 24 days and a spin-up frequency derivative of $\dot{\nu}_{2009}=(1.45\pm 0.16)\times 10^{-13}$~Hz/s. Considering the spin frequency reported in Table~\ref{tab:solution} as a proxy of the value at beginning of the 2015 outburst, we obtained $\Delta \nu_{2009-2015}=(-1.25\pm0.05)\times 10^{-6}$~Hz, which corresponds to $\dot{\nu}_{2009-2015}=(-7.2\pm0.3)\times 10^{-15}$~Hz/s. For the 2015 outburst, we found no direct evidence of frequency derivative, only an upper limit of the order of $|\dot{\nu}_{2015}|\leq 1.5\times 10^{-13}$~Hz/s ($3\sigma$ confidence level). Considering a duration of ~22 days \citep[see, Fig.~1 from][]{Papitto:2016wb}, we estimated  $-6.0\times 10^{-7}~ \textrm{Hz}<\Delta \nu_{2015-2025}<-0.4\times 10^{-7}\textrm{Hz}$, corresponding to a secular spin derivative $-1.2\times 10^{-15}~\textrm{Hz/s}< \dot{\nu}_{2015-2025} <-0.1\times 10^{-15}~\textrm{Hz/s}$. It remains unclear what causes the discrepancy between the two estimates of spin derivatives during the quiescence phases. However, the poor knowledge of the possible accretion torque during the 2015 outburst in terms of strength and duration does not allow a better constraint.    

Nonetheless, we note that similar values of spin-down derivative during the quiescence phase have been reported for a handful of AMXPs, such as SAX J1808.4$-$3658 \citep[$\dot{\nu}_{sd}=-1.152(56)\times 10^{-15}$~Hz/s; ][]{Illiano:2023aa}, IGR J17498$−$2921 \citep[$\dot{\nu}_{sd}=-4.1(2)\times 10^{-15}$~Hz/s; ][]{Illiano:2024aa}, XTE J1751$−$305 \citep[$\dot{\nu}_{sd}=-5.5(1.2)\times 10^{-15}$~Hz/s; ][]{Riggio:2011vv}, IGR J00291$+$5934 \citep[$\dot{\nu}_{sd}=-4.1(1.2)\times 10^{-15}$~Hz/s; ][]{Patruno:2010tm, Papitto:2011uv, Hartman2011a}, SWIFT J1756.9$−$2508 \citep[$\dot{\nu}_{sd}=-4.8(6)\times 10^{-16}$~Hz/s; ][]{Sanna:2018aa}.

Interpreting $\dot{\nu}_{2009-2015}$ as the spin derivative during the rotation-powered phase of \igr{} in quiescence, we can combine the rotational-energy loss rate with the rotating magnetic dipole emission to infer the magnetic field strength at the NS polar caps. For a rotating dipole in the presence of matter, we can approximate the dipolar magnetic moment of the NS as
\begin{equation}
\mu \simeq 2.1\times10^{26}\left(\frac{1}{1+\sin^2{\alpha}}\right)^{-1/2} I_{45}^{1/2}\nu_{245}^{-3/2}|\dot{\nu}|_{-15}^{1/2}\,\,\,\, \text{G cm}^3,
\label{eq:mag}
\end{equation}
where $\alpha$ is the angle between the rotation and magnetic axes \citep[see, e.g.,][and references therein]{Spitkovsky:2006uz}, $I_{45}$ is the moment of inertia of the NS in units of $10^{45}$~g~cm$^2$, $\nu_{245}$ and $|\dot{\nu}|_{-15}$ are in units of \igr{} spin frequency, and spin frequency derivative, respectively. 
Without prior knowledge of the angle $\alpha$, we can constrain the magnetic moment $\mu$ between the extremes of $\alpha=0$~deg (aligned rotator) and $\alpha=90$~deg (orthogonal rotator). Substituting the relevant
spin frequency and its secular derivative in Eq.~\ref{eq:mag}, we can limit the NS magnetic moment to be $5.6\times10^{26}\,\,\, \textrm{G\,cm$^3$}<\mu < 7.9\times10^{26}\,\,\, \textrm{G\,cm$^3$}$.
Adopting the FPS equation of state for a 1.4~M$_\odot{}$ NS and $R_{NS}=1.14\times10^{6}$~cm \citep[NS, see e.g.,][]{Friedman1981a, Pandharipande1989a}, we can constrain the magnetic field strength at the magnetic polar caps, $B_{PC}= 2 \mu/R_{NS}^{3}$, to the interval $7.6\times 10^{8}\,\,\, \textrm{G}<B_{PC}<1.1\times 10^{9}~\textrm{G}$, in accordance with the estimates from disc and stellar magnetic field interaction reported by \citet{Mukherjee:2015td}. %
However, the spin-up rate of the order of $(1.45 \pm 0.16)\times 10^{−13}$~Hz/s reported by \citet{Riggio:2011vs} from the 2009 outburst, suggests we are estimating only a lower limit on the magnetic field strength of the source, since we are ignoring the likely increase of the spin frequency of \igr{} during outburst phases where it is expected to be increasing due the matter and angular momentum transferred onto its surface.

\subsection{Secular orbital period evolution}

\label{sec:orb_ev}

We started by collecting the most accurate estimates of the orbital period from previous outbursts to investigate the secular evolution of the orbital period. More specifically, from Table~\ref{tab:solution} we considered $P_\mathrm{orb,2009}=12487.5118(4)$~s, $P_\mathrm{orb,2015}=12487.507(1)$~s and $P_\mathrm{orb,2025}=12487.505(2)$~s. We modelled the orbital evolution with a linear function from which we obtained an estimate of the period derivative of $\dot{P}_\mathrm{orb}=-1.81(67)\times 10^{-11}$~s/s with a reduced $\chi^2_\mathrm{red}=3.5$ for one dof. The relatively large uncertainty does not allow us to determine the orbital evolution unambiguously, although it seems to point towards a fast contraction on long timescales.

We explored the possibility of improving the constraint on the orbital period derivative by applying a coherent-orbital-phase analysis using the measurements of the epoch of passage of the ascending node ($T_\mathrm{ASC}$) in the three observed outbursts. More specifically, from Table~\ref{tab:solution} we considered $T_\mathrm{ASC,2009}=55088.0320286(6)$~MJD, $T_\mathrm{ASC,2015}=57107.858808(1)$~MJD, and $T_\mathrm{ASC,2025}=60717.674090(3)$~MJD. Moreover, to be able to determine the number of elapsed orbital cycles (N) and calculate the delays with respect to the predicted times $T_\mathrm{ASC, PRE}(N)=T_\mathrm{{ASC,0}}+N\, P_\mathrm{{orb,0}}$ \citep[see, e.g.,][]{Papitto:2005vm, di-Salvo:2008uu, Hartman:2009tq, Burderi:2009td, Iaria:2015vw}, we defined $P_\mathrm{orb,0}=P_\mathrm{orb,2009}$. Following the method described in \citet{Sanna:2018aa}, we find that the number of elapsed cycles between the outbursts is unambiguously determined, verifying the condition required to apply the timing technique.
 
For each outburst we determine the quantity $\Delta T_\mathrm{ASC}=T_\mathrm{ASC, OBS}-T_\mathrm{ASC, PRE}$ and modelled it using the quadratic function: 
\begin{eqnarray}
\label{eq:fit_tstar}
\Delta T_\mathrm{ASC} = \delta T_\mathrm{ASC,0} + N\, \delta P_\mathrm{orb,0}+0.5\,N^2\, \dot{P}_\mathrm{orb}P_\mathrm{orb,0}
\end{eqnarray} 
where $\delta T_\mathrm{ASC,0}$ represents the correction to $T_\mathrm{ASC,2009}$ adopted as a reference value, $\delta P_\mathrm{orb,0}$ is the correction to the orbital period, and $\dot{P}_\mathrm{orb}$ is the orbital-period derivative. 

\begin{figure}
\centering
\includegraphics[width=\columnwidth]{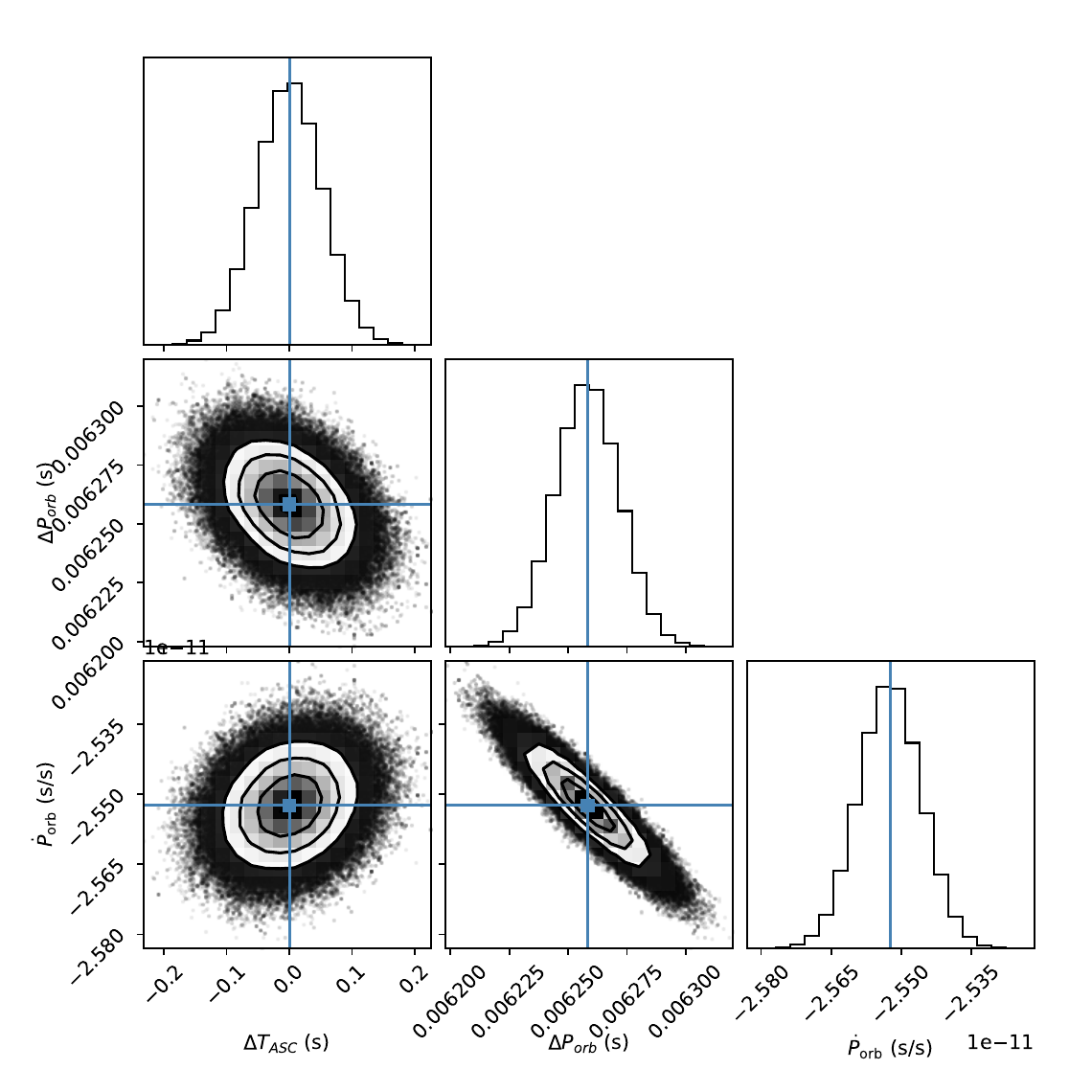}
\caption{Corner plot summarising the posterior probability distributions and pairwise covariances of the parameters $\Delta T_\mathrm{ASC}$, $\Delta P_\mathrm{orb}$, and $\dot{P}_\mathrm{orb}$ obtained from the Bayesian MCMC analysis of the secular evolution of the $T_\mathrm{ASC}$ from the three outbursts of \igr{}, see the text in Sec.~\ref{sec:orb_ev} for more details. Solid lines delimit the $1\sigma$ and $2\sigma$ confidence level regions.}
\label{fig:corner_plot}
\end{figure} 

Because we only have three data points and three parameters in our quadratic model, a standard least-squares approach would yield no dof and thus no meaningful uncertainty estimates. Instead, we employed a Bayesian method with suitable priors to constrain the parameter space and obtain reliable intervals from the resulting posterior distributions. We estimated the best-fit parameters and their uncertainties by sampling the posterior distribution with a Markov Chain Monte Carlo (MCMC) algorithm. Specifically, we adopted flat (uniform) priors for the parameters $\delta T_\mathrm{ASC,0}$ and $\delta P_\mathrm{orb,0}$ based on the accuracy of the available measurements. Values significantly larger than the two quantities would have led to clear observational signatures during the 2009 outburst, thereby justifying our chosen prior bounds. We employed the \textit{emcee} MCMC ensemble sampler \citep{Foreman-Mackey:2013aa} to generate sufficient draws from the joint posterior. We then derived the following median parameter values and their $1\sigma$ intervals from the 16th and 84th percentiles of the resulting marginalized posterior distributions: $\delta T_\mathrm{ASC,0} = <0.15 ~s$ (3$\sigma$~c.l.), $\delta P_\mathrm{orb,0}= 6.26(1)\times 10^{-3}$~s, $\dot{P}_\mathrm{orb}= -2.55(7)\times 10^{-11}$~s/s. This result implies a more accurate estimate of the orbital period of $P_\mathrm{orb}=12487.51806(2)$~s. It is worth noting that the orbital period derivative is compatible with the estimate obtained from the direct comparison of the orbital periods. However, given the accuracy achieved in measuring the orbital period during the 2009 outburst, we should have been able to directly measure the orbital period obtained from the $T_\mathrm{ASC}$ secular evolution. This crucial point needs further investigation to understand at which level secular changes in $T_\mathrm{ASC}$ trace back the orbital evolution.

Erratic orbital period variation has been reported for several binary pulsars classified as MSPs. At least three black widow pulsars (characterised by a $<0.1$~M$_\odot$ low-mass, partially evaporated or oblate, companion star) exhibited complex orbital changes requiring high-order orbit derivatives to be described. Among them, PSR J0023+0923 interestingly displays changes in the orbital period, measured in terms of large erratic drifts of the time of the predicted $T_{\text{ASC}}$ both in radio and X-rays \citep[see, e.g., ][]{Nielsen:2020aa}. Similar behaviour has been reported for red back pulsars (with relatively more massive, 0.1-0.4~M$_\odot$, non-degenerate companions). Systems like PSR J2339−0533 show alternating increases and decreases in the orbital period over timescales of a few years. Cyclic changes in the companion star’s gravitational quadrupole moment, likely associated with magnetic activity cycles, have been proposed to explain the peculiar orbital behaviour \citep[see, e.g.,][and references therein]{Pletsch2015a}. The transitional MSP PSR J1023+0038 shows erratic behaviour of the time of passage at the ascending node, with variations of tens of seconds over timescales of years \citep[see, e.g.,][]{Archibald2013a, Archibald:2015vw, Jaodand:2016aa, Papitto:2019aa, Burtovoi:2020aa, Illiano:2023ab}. Long-term timing studies revealed significant timing noise and orbital variations for red backs in globular clusters. For example, orbital changes in MSPs located in Terzan 5 have been interpreted with a combination of cluster-induced accelerations and intrinsic processes like mass transfer or torques from the companion star \citep[see, e.g.,][]{Rosenthal:2025aa}.

Currently, significant orbital period derivatives have been measured for three other AMXPs, i.e., SAX J1808.4$-$3658 \citep[see e.g.,][]{di-Salvo:2008uu, Patruno:2017ah, Sanna:2017vj, Bult:2020tu, Illiano:2023aa}, SWIFT J1749.4$-$2807 \citep[][]{Sanna:2022tt} and IGR J17062$-$6143 \citep{Bult:2021vs}, all suggesting a fast orbital expansion. For five more AMXPs, uncertainties are such that it is not yet possible to determine whether the orbits are secularly expanding or shrinking \citep[see, e.g.,][]{Sanna:2016ty, Bult:2018ve, Sanna:2018aa, Sanna:2018tx, Patruno:2017vp, Sanna:2017tx}.

Considering all the previously discussed caveats, we notice that both independent methods investigated to extrapolate the secular orbital evolution suggest a fast contraction of the binary system. If confirmed, this could be compatible with a scenario in which angular momentum loss due to gravitational wave radiation dominates. Although theoretically predicted, this scenario has never been observationally confirmed for AMXPs so far. If we define the orbital evolution timescale as $\tau = P_{orb}/\dot{P}_{orb}$, and we consider an orbital period derivative of the order of $\dot{P}_{orb}\simeq -2\times 10^{-11}$~s/s, we find that \igr{} is evolving with $\tau \sim20$~Myr. For comparison, let us consider the evolutionary timescale reported by \citet{Paczynski71}:
\begin{equation}
    \tau_{gw}=380\frac{(1+q)^2}{q}\left(m_1+m_2\right)^{-5/3}(P_{orb, days})^{8/3} \text{Gyr},
\end{equation}
where $q=m_2/m_1$ is the binary mass ratio, $m_1$ and $m_2$ are the NS and the secondary masses in units of solar masses, and $P_{orb, days}$ is the orbital period expressed in days. Adopting $m_2$ in the range 0.15-0.44~M$_\odot$ \citep[see, e.g.,][]{Papitto:2010aa}, $m_1$ in the range 1.0-2.5~M$_\odot$, we find a predicted evolutionary timescale for \igr{} in the range $\tau_{gw}=3-15$~Gyr, almost three orders of magnitude longer than our estimate. 

Following \citet{di-Salvo:2008uu}, we can express the orbital period derivative of the system in terms of the angular momentum lost and transferred in the system. More specifically, if we express the orbital angular momentum of the system as: $J_{\rm orb} = [G a / (M_1 + M_2)]^{1/2} M_1 M_2$ (where $G$ is the Gravitational constant, and $a$ is the orbital separation), and define $- \dot M_2$ the mass transferred by the secondary (which can be accreted 
onto the NS, conservative mass transfer,  or can be lost from the system, non-conservative mass transfer), we can write the expression:
\begin{equation}
\label{eq:dotP}
\frac{\dot P_{\rm orb}}{P_{\rm orb}} = 3 \left[\frac{\dot J}{J_{\rm orb}} - 
\frac{\dot M_2}{M_2} \; g(\beta,q,\alpha)\right],
\end{equation}
where $g(\beta,q,\alpha) = 1 - \beta q - (1-\beta) (\alpha + q/3)/(1+q)$, with $\beta$ the fraction of mass lost by the secondary and accreted by the NS, $\alpha$ the specific angular momentum of the mass lost by the system, and
$\dot J / J_{\rm orb}$ the possible losses of angular momentum from the system. 

According to theory, orbital angular momentum can be dissipated via gravitational waves and magnetic braking emissions. The latter being dominant for systems with longer orbital periods \citep[starting from $P_{orb}>2$~h; see, e.g.,][]{Verbunt:1993vj, Tauris:2001aa}. Combining the two mechanisms, we can express the angular momentum loss as:
\begin{equation} 
\label{eq:GRMB}
\left(\frac{\dot{J}}{J_{\rm orb}}\right)_{\rm GR+MB}=-2.1\times 10^{-17}m_1~m_{2}~m^{-1/3}P_{\rm orb,3h}^{-8/3}[1+T_{\rm MB}],
\end{equation}
where $m_1$, $m_2$, and $m$ represent the primary, secondary, and total mass in units of solar masses. $P_{\rm orb,3h}$ is the orbital period in units of three hours and $T_{\rm MB}$ represents the magnetic braking torque. Following \citet[][]{Iaria:2018tq}, $T_{\rm MB}$ can be parameterized as:
\begin{equation}
\label{eq:TMB}	
T_{\rm MB}=18.9(k^2)_{0.1}f^{-2}m_1^{-1}P_{\rm orb,3h}^2\left(\frac{m_2}{m_1}\right)^{1/3}\left(1+\frac{m_2}{m_1}\right)^{2/3},
\end{equation}
where $k$ represents the gyration radius of the secondary star and $f$ is a dimensionless parameter which can be considered either as $f=0.73$ \citep{Skumanich:1972vy} or $f=1.78$ \citep{Smith:1979vn}.

According to \citet{Papitto:2010aa}, the companion star can be limited in the interval $M_2/M_\odot=0.15-0.44$ assuming $M_1=1.4$~M$_\odot$. Assuming $k=0.45$ \citep[see. e.g.,][]{Wadhwa:2024aa}, and $f=0.73$ (which maximises the magnetic braking effect), Equation~\ref{eq:GRMB} applied to \igr{} limits $\dot{J} / J$ between $-9\times 10^{-17}$~s$^{-1}$ ($m_2=0.15$) and $-4\times 10^{-16}$~s$^{-1}$ ($m_2=0.44$), which according to Equation~\ref{eq:dotP}, could account up to $\sim 5$\% or $\sim 20$\% of the observed $\dot P_{\rm orb}/P_{\rm orb}$, respectively.

Recent developments, including the convection and rotation-boosted (CARB) magnetic braking model proposed by \citet{Van:2019aa}, offer more physically motivated descriptions of angular momentum losses in low-mass, fully convective secondary stars. Unlike earlier approaches that often oversimplify or neglect convection effects, CARB predicts enhanced magnetic braking for rapidly rotating, fully convective companions. 

Applying the CARB model for a companion mass of $m_2 \sim 0.2~M_{\odot}$, with a wind mass-loss rate of $\dot{M}_w\sim 7.5\times 10^{-14}~M_{\odot}$/yr, a stellar radius $R\sim0.3R_{\odot}$ (corresponding to the secondary Roche lobe radius for the assumed companion mass), a luminosity $L\sim0.02~L_{\odot}$ \citep[parameters extrapolated from Reimers’ prescription for mass-loss rates due to stellar winds][]{Reimers:1975aa}, and assuming a rotation period of the order of the binary period due to tidally locked rotation, yields a magnetic braking torque ranging widely from approximately $-3\times 10^{46}$~g~cm$^2$/s$^2$, corresponding to a $\dot{J} / J\sim 2\times 10^{-5}$s$^{-1}$, which is orders of magnitude larger than the observed effect measured in this work. However, it is worth noting from Equation 6 in \citet{Van:2019aa}, the strong dependence of $\dot{J}_{MB}$ on the companion rotation rate $\Omega$. Releasing the tidally locked rotation condition for the companion star, and hypothetically considering rotation rates of the order of the tens of days, could provide the required amount of angular momentum losses to match the observed $\dot P_{\rm orb}/P_{\rm orb}$. Nevertheless, the very short tidal-synchronisation timescales in AMXPs make any sustained departure from synchronous (tidally locked) rotation highly improbable, rendering these slower-rotation scenarios largely hypothetical for \igr{}.  Future observational tests targeting precise measurements of orbital decay rates and companion rotation periods would be essential to differentiate between traditional MB models and the CARB scenario, ultimately refining our understanding of angular momentum evolution in AMXPs.

Let us consider now the contribution of mass transfer ($3 \dot M_2/M_2* g(\beta,q,\alpha)$) by exploring the two extreme scenarios: conservative ($\beta=1$) and totally non-conservative ($\beta=0$) mass transfer. The former implies $g(\beta,q,\alpha)=1-q$, hence, $\dot P_{\rm orb}/P_{\rm orb} \sim 3 \dot M_2/M_2* (1-q)$, ignoring the effect of $(\dot{J}/J)_{GR+MB}$. The requested rate of mass transfer from the companion would then range between $2.9\times 10^{-9}$~M$_\odot$/yr and $7.1\times 10^{-9}$~M$_\odot$/yr, considering the possible secondary mass interval. Defining $L_X = G\dot{M}M_{NS}/R_{NS}$ the luminosity emitted by the NS due to accretion of matter, we can estimate an expected constant luminosity from the NS over the long-term timescales between $3\times 10^{37}$~erg~s$^{-1}$ and $7.4\times 10^{37}$~erg~s$^{-1}$ for $M_{NS}=1.4$~M$_\odot$ and $R_{NS}=1.14\times 10^{6}$~cm. While this value is of the same order of magnitude as that estimated from spectral modelling of the source during the outburst \citep{Papitto:2010aa, Bozzo:2010aa, Papitto:2016wb}, it is not compatible with the average luminosity of \igr{} over a sixteen-year baseline from which we extrapolated the orbital evolution, making the conservative mass-transfer scenario unfeasible. It is worth noting that even accounting for the largest loss of angular momentum combining gravitational waves and magnetic braking ($\sim20$\% of the observed $\dot P_{\rm orb}/P_{\rm orb}$), the conservative mass-transfer scenario would require a long-term average luminosity still too high ($\sim 10^{37}$~erg~s$^{-1}$) to be missed during the quiescence phases. 

On the other hand, a non-conservative mass transfer scenario requires continuous mass transfer from the secondary, with only partial accretion onto the NS surface. Assuming roughly an average outburst duration of one month for three observed episodes, we could define a source duty cycle of $\sim 0.015$ over the sixteen-year baseline investigated here. Therefore, assuming that the inferred mass accretion rate during the outburst represents the average long-term mass transfer rate from the secondary, $\sim 0.015$ can also be interpreted as the fraction of accreted matter $\beta$ previously defined. As a first-order approximation, given the small value of $\beta$, we can consider the totally non-conservative mass-transfer scenario ($\beta = 0$), for which
$g(0,q,\alpha) = (1 - \alpha + 2 q / 3)/(1+q)$. Since $\dot P_{\rm orb}/P_{\rm orb}$ is negative, $g(0,q,\alpha)$ needs to be positive. Assuming the NS mass of 1.4~M$_\odot$ and exploring the companion mass in the range 0.15-0.44, we can investigate the location at which the mass needs to be ejected from the system, in terms of its specific angular momentum carried away, to produce an orbital evolution of the order of $\dot P_{\rm orb}/P_{\rm orb}$. We find that ejecting matter with an $\alpha$ parameter in the range $1.9-2.1$ solves Equation~\ref{eq:dotP}, allowing, at least theoretically, to explain the inferred fast secular contraction of the binary system with a combination of loss of mass and angular momentum from the system. However, several open questions remain: i) What causes the matter to be ejected from the secondary during the quiescence phase? If the radiation pressure is generated from a rotation-powered pulsar turning on \cite[see, e.g.,][and references therein]{di-Salvo:2008uu, Burderi:2009td}, why is there evidence of only one AMXP \citep[IGR J18245–2452][]{Papitto:2013uf} swinging between accretion and rotation-powered states? \citep[see, e.g., ][for more details on the topic]{Burderi:2009td, Iacolina:2010aa}. ii) Why is there no evidence of the ejected matter around the companion star? 

Alternative processes, such as gravitational quadrupole coupling between the companion star and the binary orbit \citep[see, e.g.,][]{Applegate:1992uh, Applegate:1994vp}, have been proposed to describe periodic orbital variations on timescales of years for different MPSs \citep{Doroshenko2001a, Pletsch2015a}. Magnetically driven changes in the mass distribution of the donor star modify the companion's oblateness, changing the gravitational interaction within the binary system and inducing orbital period variations on
dynamical timescales. Although plausible, the application of this scenario for \igr{} is currently limited by the lack of observational evidence for a long-term orbital modulation, with only three available outbursts to investigate. 

Confirming this secular trend is crucial to investigating further theoretical frameworks that explain such a peculiar orbital evolution. Therefore, future outbursts of the \igr{} are required to validate the system's orbital period derivative.

\subsection{Spectral properties}

We studied the spectral characteristics of \igr{} during its 2025 outburst using \nicer{} and \nustar{}. The spectra were well described by a semi-phenomenological model consisting of an absorbed blackbody and a thermally Comptonized component ({\tt ThComp}). \nicer{} observations revealed an average unabsorbed flux of 4$\times$10$^{-10}$ erg~s$^{-1}$~cm$^{-2}$ in the 0.5–10 keV range, which gradually increased from 3.9$\times$10$^{-10}$ to 4.2$\times$10$^{-10}$ergs$^{-1}$cm$^{-2}$ between February 11 and 14, 2025. The source flux declined to 2.52$\times$10$^{-10}$ erg~s$^{-1}$~cm$^{-2}$ in the same energy range as observed by \nustar{} on February 19–20, 2025, with a corresponding bolometric flux (0.5–100 keV) of 6.7$\times$10$^{-10}$ erg~s$^{-1}$~cm$^{-2}$ or a luminosity of 3.8$\times$10$^{36}$ erg~s$^{-1}$~cm$^{-2}$ at a distance of 6.9~kpc. This flux decline suggests that the \nustar{} observation occurred during the decay phase of the 2025 outburst.  In comparison, during the 2009 \citep{Bozzo:2010aa} and 2015 \citep{Papitto:2015wo} outbursts, the peak flux in the 0.5–10 keV band was measured in the range of (5–7)$\times$10$^{-10}$ erg~s$^{-1}$~cm$^{-2}$, slightly higher than the flux observed by \nicer{} in 2025. Assuming similar outburst behaviour, the outburst likely peaked between the observation gap of NICER and NuSTAR (see also \citealt{2025ATel17061....1S}).

The spectral shape of \igr{} is dominated by a hard power-law component with an index around 1.6–1.8 and an electron temperature of $\ge$15~keV. We also observed an emission line feature at 1.7 keV with NICER, consistent with the K$\alpha$ transition of neutral~Si. The same feature has been observed in other \nicer{} observations \citep[see, e.g.,][]{Marino:2022wx, Illiano:2024aa}, suggesting an instrumental origin. Additionally, both instruments detected weak emission residuals around 6.6$\pm$0.2 keV, with a central line energy consistent with the K$\alpha$ transition of neutral or He-like Fe. This line was relatively broader, with a width of 0.6$\pm$0.2 keV. \xmm{} also observed an iron line at  6.65$\pm$0.25~keV with a relatively broad width of >0.56~keV during the decay phase of the 2015 outburst, with an inner accretion disc radius truncated at $>$40~km \citep{Papitto:2016wb}.
To examine the emission from the accretion disc and the origin of the 6.6 keV line, we modelled the \nustar{} spectrum with the self-consistent reflection model {\tt relxillCp}. This allowed us to constrain the inclination angle of the system to be greater than $37^\circ$, which aligns with earlier measurements from the 2015 outburst \citep{Papitto:2016wb}. 
The modelling also indicated that the inner disc radius is truncated in the range of (40–180)~$\rm R_g$, corresponding to (82–370)~km, inferred assuming a canonical NS mass of 1.4 solar mass. A slightly better constrained value was found in the range of (24–122)~$\rm R_g$, corresponding to (50–252)~km from broadband fitting using {\tt Diskline}. The lower limit of inner disk radius is almost closer to the predicted NS co-rotation radius of $\sim43$~km.
We found a moderate ionisation of the accretion disc, with a log$\xi$ of 2.7$\pm$0.7, supporting an ionised iron emission line from the inner region of the accretion disc. The line broadness can be explained by considering the effect of the Doppler shift and the NS gravitational field.

\subsection{Type-I X-ray burst}

A thermonuclear X-ray burst was observed during the \nustar{} observation of \igr{}. The burst exhibited a rapid rise time of 2~s, followed by an exponential decay of 8.5~s, with a total duration of 25~s. Typically, bursts fueled by hydrogen-rich material are longer, lasting up to hundreds of seconds \citep{Bildsten2000AIPC..522..359B}. The short duration of this burst suggests a fuel composition deficient in hydrogen. However, it could also imply a higher CNO metallicity, as noted by \citet{2011A&A...529A..68F}, based on the observed burst recurrence time of 7 hours in the 2009 outburst. The NuSTAR burst occurred at a pre-burst flux of (6.62$\pm$0.05)$\times$10$^{-10}$~erg~s$^{-1}$cm$^{-2}$, corresponding to a luminosity of 3.8$\times$10$^{36}$ erg~s$^{-1}$ at a distance of 6.9~kpc. This suggests that the NS was accreting at the mass accretion rate of about 1\% of the Eddington limit \citep{Kuulkers03}. Furthermore, our timing studies revealed oscillations during the rise and decay phases of the \nustar{} burst around 244.85 Hz, with a significance level of 3.6$\sigma$ and a fractional amplitude of 25$\pm$5\%.  

Our burst time-resolved spectroscopy revealed a peak blackbody temperature of $\sim$3~keV, cooling to 1.2~keV by the end of the burst. The blackbody normalisation at the peak temperature corresponds to an emission radius of 5.4$\pm$0.5~km. During the decay phase, a higher emission radius of 8.7$\pm$0.6~km, consistent with the NS radius, was observed. The burst lacked evidence of photospheric radius expansion, indicating that the observed peak flux of (5.4$\pm$0.4)$\times$10$^{-8}$ erg~s$^{-1}$~cm$^{-2}$ was below the Eddington limit. This peak value is consistent with non-PRE bursts previously observed with \rxte{} and \inte/JEM-X, which ranged between (3–6.7)$\times$10$^{-8}$erg~s$^{-1}$~cm$^{-2}$ and provided an upper limit of the distance \citep{Altamirano:2010aa, 2011A&A...529A..68F}. The fluence of the NuSTAR burst was estimated to be (5.1$\pm$0.1)$\times$10$^{-7}$erg~cm$^{-2}$, corresponding to a net energy release of 2.9$\times$10$^{39}$~erg. No evidence of soft excesses was detected during the burst.

\begin{acknowledgements}
A.P. is supported by INAF (Research Grant ‘Uncovering the optical beat of the fastest magnetised NSs 620 (FANS)’), the Italian Ministry of University and Research (MUR) PRIN 2020 Grant 2020BRP57Z, ‘Gravitational and Electromagnetic-wave Sources in the Universe with current and next-generation detectors (GEMS)’) and Fondazione Cariplo/Cassa Depositi e Prestiti, grant no. 2023-2560. M.N. is a Fonds de Recherche du Québec – Nature et Technologies (FRQNT) postdoctoral fellow.
C.M. is supported by INAF (Research Grant `Uncovering the optical beat of the fastest magnetised neutron stars 620 (FANS)') and the Italian Ministry of University and Research (MUR) (PRIN 2020, Grant 2020BRP57Z, `Gravitational and Electromagnetic-wave Sources in the Universe with current and next-generation detectors (GEMS)').
T.B. is supported in part by the Turkish Republic, Directorate of Presidential Strategy and Budget project, 2016K121370. 
\end{acknowledgements}

\bibliographystyle{aa} 
\bibliography{biblio.bib}

\begin{appendix}
\section{MCMC results from the reflection modelling}
\begin{figure}[!h]
\centering
\includegraphics[width=0.8\textwidth]{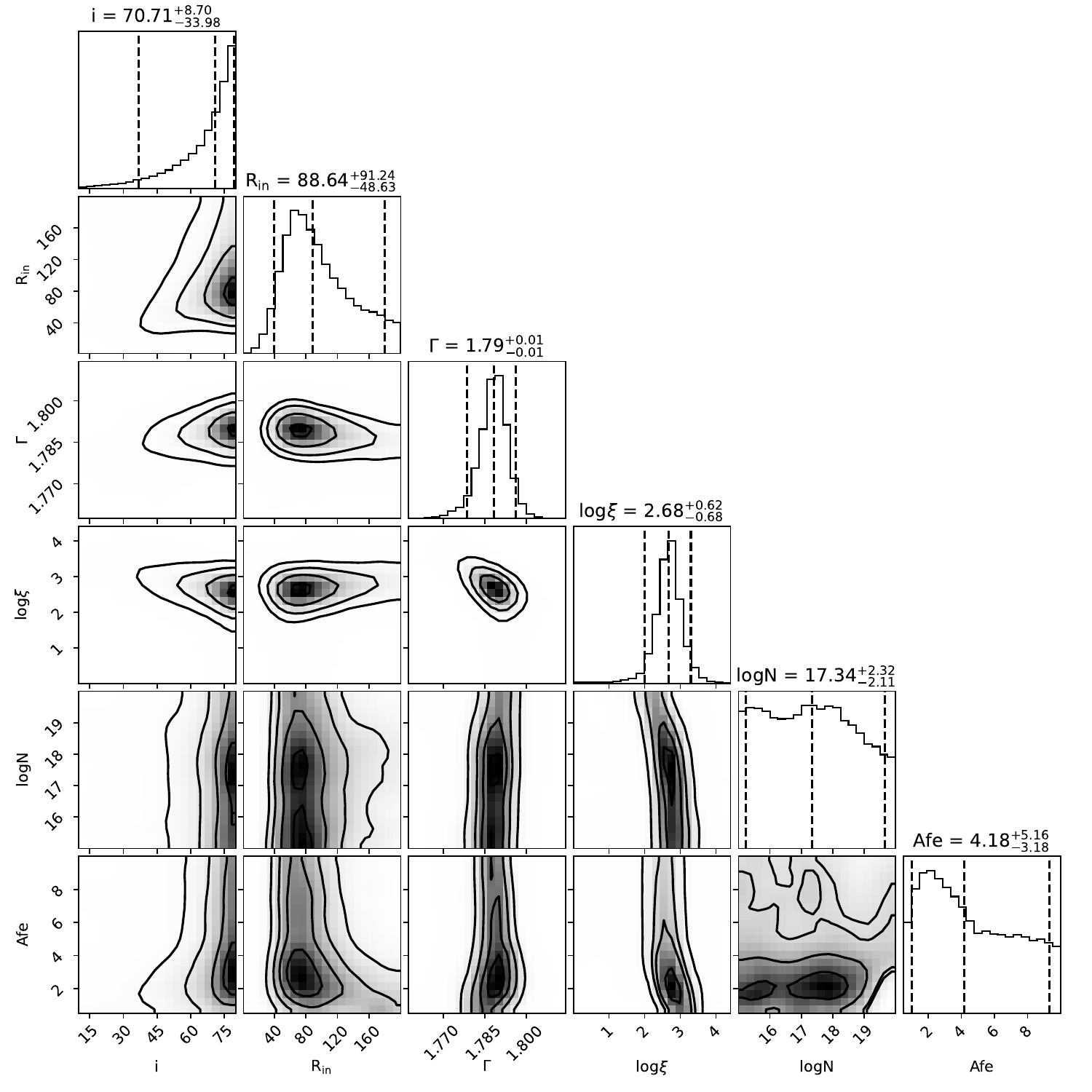}
\caption{MCMC results from the reflection modelling of  NuSTAR data. The modelling parameters are inclination angle (i), inner disc radius ($\rm R_{in}$) in units of gravitational radii,  photon index ($\Gamma$), the accretion disc's ionisation parameter (log$\xi$), disc density (log~N),  and iron abundance relative to solar (Afe).   The uncertainties on parameters are given for a 90\% confidence range. }
\label{fig:corner_plot}
\end{figure} 

\end{appendix}

\end{document}